\documentclass[10pt]{article}
\usepackage[letterpaper,hmargin=1in,vmargin=1in]{geometry}
\usepackage{times}
\usepackage{tikz}
\usepackage[utf8]{inputenc}
\usepackage{amsmath,amssymb}
\usepackage{subfigure}
\usepackage{bm}
\usepackage{mathtools}
\usepackage{enumitem}
\usepackage[hidelinks]{hyperref}
\usepackage[capitalise]{cleveref}
\usepackage{bbold}
\usepackage{color}
\usepackage{soul}
\usepackage{lettrine}
\usepackage{placeins}
\usepackage{titlesec}

\usetikzlibrary{matrix,decorations.pathreplacing,calc,fit,positioning, arrows.meta,calligraphy}
\usepackage{hf-tikz}

\newcommand\address[1]{\hskip2.25pc \parbox{.8\textwidth}{ \noindent%
   \footnotesize \it \begin{center} #1 \end{center}\rm }  \normalsize \vskip-.2cm }

\renewcommand\title[1]{\bf \hskip2.25pc \parbox{.8\textwidth}{ \noindent%
   \LARGE \bf \begin{center} #1 \end{center} \rm } \vskip.1in \rm\normalsize }

\renewcommand\author[1]{\hskip2.25pc \parbox{.8\textwidth}{ \noindent%
   \normalsize \bf \begin{center} #1 \end{center}\rm } \vskip-1pc }
\makeatletter

\newcommand{\ket}[1]{\vert \,#1 \,\rangle}
\newcommand{\bra}[1]{\langle \, #1\,\vert}
\newcommand{\E}[1]{\langle \, #1 \, \rangle}

\titleformat{\subsection}[runin]
{\normalfont\bfseries}
{}{0pt}{}[]
\titlespacing{\section}
{\parindent}{1.5ex plus .1ex minus .2ex}{0pt}

\titleformat{\section}{\Large\normalfont\bfseries}{}{0pt}{}
\titlespacing*{\section}{0cm}{0.3cm}{0.2cm}

\begin{document}

\title{Quantum Optimisation for Continuous Multivariable Functions by a Structured Search}

\author{Edric Matwiejew$^{\dagger}$, Jason Pye, Jingbo B. Wang}

\address{
$^{1}$ Department of Physics, The University of Western Australia, Perth, WA 6009, Australia\\
$^{\dagger}$edric.matwiejew@uwa.edu.au}

\begin{quote}
\noindent\textbf{Solving optimisation problems is a promising near-term application of quantum computers. Quantum variational algorithms leverage quantum superposition and entanglement to optimise over exponentially large solution spaces using an alternating sequence of classically tunable unitaries. However, prior work has primarily addressed discrete optimisation problems. In addition, these algorithms have been  designed generally under the assumption of an unstructured solution space, which constrains their speedup to the theoretical limits for the unstructured Grover's quantum search algorithm. In this paper, we show that quantum variational algorithms can efficiently optimise continuous multivariable functions by exploiting general structural properties of a discretised continuous solution space with a convergence that exceeds the limits of an unstructured quantum search. We introduce the Quantum Multivariable Optimisation Algorithm (QMOA) and demonstrate its advantage over pre-existing methods, particularly when optimising high-dimensional and oscillatory functions.}
\end{quote}

\lettrine[lraise=0.2]{Q}{uantum} computing promises to solve problems that are classically intractable~\cite{matthews_how_2021}. One potential application is optimisation over high-dimensional spaces, which suffers from the long-fought `curse of dimensionality'~\cite{bellman_dynamic_1957}. Quantum computers may help overcome this by leveraging quantum superposition and entanglement on exponentially large solution spaces. For this reason, much attention has been applied to the development of quantum optimisation algorithms~\cite{farhi_quantum_2014, hadfield_quantum_2019, marsh_combinatorial_2020,guerreschi_practical_2017}. 
Quantum Variational Algorithms (QVAs) belong to a class of hybrid quantum-classical algorithms in which a classically parameterised quantum system accelerates a search through a finite problem solution space~\cite{farhi_quantum_2014,marsh_quantum_2019,marsh_combinatorial_2020,matwiejew_quop_mpi_2022}. These algorithms have a flexible structure and are inherently resilient to error~\cite{peruzzo_variational_2014,cerezo_variational_2021}. As such, they are predicted to be an early practical use of quantum computing in the Noisy Intermediate-Scale Quantum (NISQ) era~\cite{preskill_quantum_2018}.

QVA development has primarily focused on Combinatorial Optimisation Problems (COPs).
These include algorithms for unconstrained optimisation, such as the widely studied Quantum Approximate Optimisation Algorithm (QAOA), and constrained optimisation~\cite{farhi_quantum_2014,marsh_quantum_2019,marsh_combinatorial_2020,hadfield_quantum_2019}. Most COPs of practical concern lack identifiable structure~\cite{marsh_quantum_2019,marsh_combinatorial_2020}. Consequently, QVAs for COPs strive for an unbiased search over the space of valid solutions. In this regard, they are fundamentally related to Grover's search algorithm~\cite{grover_fast_1996}, a well-known quantum algorithm for deterministic search in an unstructured solution space. Grover's search is optimal for the number of required solution space evaluations and thus sets an upper bound on the efficiency of QVAs for general COPs~\cite{zalka_grovers_1999,bennett_quantum_2022}. The utility of QVAs for COPs arises from the reality that Grover's search requires a quantum circuit depth that is not feasible on near-term hardware. However, as QVAs utilising an unstructured search can provide a sub-Grover speedup at best, there is motivation to develop algorithms capable of exceeding this limit by exploiting solution space structure. One example is the QAOA for the case of the max-cut problem on three-regular graphs~\cite{wurtz_maxcut_2021}. There is also recent work adopting an iterative process to construct general problem-tailored QAOA operators~\cite{zhu_adaptive_2022}.

Structured approaches are ubiquitous in classical algorithms for Continuous Multivariable Optimisation Problems (CMOPs) with many well-known instances, including the gradient-based \discretionary{Broyden-Fletcher-}{Goldfarb–Shanno}{Broyden-Fletcher-Goldfarb–Shanno} (BFGS) algorithm and the simplex-based Nelder-Mead algorithm~\cite{nocedal_numerical_2006,nelder_simplex_1965}. However, QVAs for CMOPs have received little attention despite their relevance to optimisation tasks considered in the literature on QVAs for COPs. For example, the financial portfolio optimisation problem typically includes proportion-weighted asset combinations---which are not accounted for by a combinatorial approach~\cite{markowitz_portfolio_1952,slate_quantum_2021}.
Various quantum algorithms for gradient descent have been developed outside of the QVA framework. Some approaches are based on an algorithm of Jordan~\cite{jordan_fast_2005}, which provides a speedup in the computation of gradients in high-dimensional spaces. Measurements of this gradient are then used in an iteration of a descent algorithm~\cite{jordan_fast_2005,gilyen_optimizing_2019,chakrabarti_quantum_2020,van_apeldoorn_convex_2020,zhang_quantum_2021}. Another approach is to use amplitude encoding of solution vectors and leverage quantum speedups in the solution of linear systems~\cite{rebentrost_quantum_2019,kerenidis_quantum_2020}.
A gradient-descent-inspired QVA for continuous-variable optimisation was suggested in \cite{verdon_quantum_2019}. The authors numerically demonstrated a wavepacket propagating towards the global minimum of a two-dimensional function with hand-selected variational parameters~\cite{verdon_quantum_2019}. A recent experimental implementation is described in~\cite{enomoto_continuous-variable_2022}. This is referred to here as Quantum Optimisation with Wavepacket Evolution (QOWE).

Building on this, we have developed the highly efficient Quantum Multivariable Optimisation Algorithm (QMOA), which adopts a continuous-time quantum walk (CTQW) framework~\cite{QWbook2014}. It solves CMOPs by implementing semi-independent CTQWs on a composite graph structure which conforms to the structure of the discretised solution space. Using a circulant operator structure with the quantum Fourier transform makes the QMOA efficient~\cite{qiang2016, loke2017, zhou2017}, while supporting independent parameterisation over the solution space dimensions.

\section{Results}\label{sec:results}

\noindent This section introduces the main contribution of this work, the QMOA, as an extension of the QAOA and QOWE. We begin by introducing a quantum encoding of the continuous-variable optimisation problem and the general form of a QVA. The QMOA is then developed by considering the relative ability of mixing operators in the QAOA and QOWE to capture structural information and distinguish between unique solutions in the quantum-encoded solution space. We present numerical results that assess these QVAs in terms of mean error, statistical distance from the global minimum and maximum amplification. To identify efficient exploitation of solution space structure, we compare QVA maximum amplification to the amplification produced by a deterministic restricted depth Grover's search (RDGS) (see App. \ref{app:grovers}). For the two best-performing QVAs, the QMOA with a complete graph mixer and the QAOA with a hypercube mixer, we consider the mean error and statistical distance over twenty test-functions (see App. \ref{app:test_functions}) and empirically assess their scalability in terms of the function dimension and grid size in each dimension.

\subsection{The Continuous Multivariable Optimisation Problem.} For a continuous function $f: X \rightarrow Y$, where $X \subset \mathbb{R}^D$ and $Y \subset \mathbb{R}$, continuous-variable optimisation seeks ${\bm{x}^*} = (x_0, ..., x_{D-1}) \in X$ satisfying,  
\begin{equation}
	f({\bm{x}^*}) \leq f_\text{min} + \epsilon,
	\label{eq:optimisation}
\end{equation}
where $f_\text{min}$ is the global minimum of $f$ and $\epsilon > 0$ defines a region of accepted $\bm{x}$ near $f_\text{min}$.

An encoding of the optimisation problem in a system of qubits consists of evaluating $f$ on a grid of $K = N^D$ points. In each dimension $d$, the coordinate is discretised as $x_{d,n_d} = x_{d,0} + n_d \Delta x_d$, with minimum value $x_{d,0}$, grid spacing $\Delta x_d$, and $n_d = 0, \dots, N-1$. The complete solution space of discretised coordinates $\bm{x}_k$ is then represented using $\mathcal{O}(\log K)$ qubits by states $\ket{k} \equiv \ket{x_{D-1,n_{D-1}},x_{D-2,n_{D-2}}, \dots, x_{0,n_0}}$, where $k = 0, \dots, N^D-1$ is a vectorised index for the set $(n_0, \dots, n_{D-2}, n_{D-1}) \in \{ 0, \dots, N-1 \}^D$. For optimisation over this discrete space, we denote the global minimum as ${\bm{x}^*} \equiv \text{argmin}_{k} f(\bm{x}_k)$.

\vspace{0.3cm}\noindent\textbf{QVAs for Approximate Optimisation.} This work considers QVAs of the form 
\begin{equation}
	\label{eq:variational_generic}
	\ket{\bm{t},\bm{\gamma}}=\left( \prod_{i = 1}^{p}\hat{U}(t_i,\gamma_i) \right)\ket{\psi_0},
\end{equation}
where the positive integer $p$ is a fixed number of ansatz iterations, $\hat{U}$ is the ansatz unitary, $t_i$ and $\gamma_i$ are real-valued variational parameters and, 
\begin{equation}
	\ket{\psi_0} = \frac{1}{\sqrt{K}}\sum_{k = 0}^{K-1}\ket{k},
	\label{eq:equal_superposition}
\end{equation}
unless otherwise specified.
The ansatz unitary consists of the so-called alternating phase-shift, $\hat{U}_Q$, and mixing, $\hat{U}_W$, unitaries,
\begin{equation}
	\hat{U}(t,\gamma)=\hat{U}_W(t)\hat{U}_Q(\gamma).
\end{equation}
The first of these, 
\begin{equation}
	\hat{U}_Q(\gamma) = \exp(-\text{i} \gamma \hat{Q}),
	\label{eq:phase-shift}
\end{equation}
applies a phase-shift proportional to
\begin{equation}
	\hat{Q} = \sum_{k=0}^{K - 1} f_k \ket{k}\bra{k},
\end{equation}
where $f_k \equiv f(\bm{x}_k)$. The second unitary, $\hat{U}_W$, conforms to some structure specific to each algorithm. Its role is to drive the transfer of probability amplitudes between the solution states. During the mixing stage, phase differences encoded by $\hat{U}_Q$ result in interference that is manipulated by varying $(\bm{t}, \bm{\gamma})$.

A QVA then proceeds by repeated preparation of $\ket{\bm{t}, \bm{\gamma}}$. After each iteration, $(\bm{t}, \bm{\gamma})$ are tuned using a classical optimisation algorithm to minimise the expectation value
\begin{equation}
	\langle Q \rangle = \bra{\bm{t}, \bm{\gamma}} \,\hat{Q}\, \ket{\bm{t}, \bm{\gamma}}.
\end{equation}
The intended consequence is an increased probability of measuring solutions satisfying \cref{eq:optimisation}. The possible amplification increases with $p$ at the expense of a deeper quantum circuit and larger parameter space for the classical optimiser.

\subsection{The Quantum Approximate Optimisation Algorithm.} The QAOA defines the mixing unitary as
\begin{equation}
	\hat{U}_{W\text{-QAOA}}(t) = \exp(-\text{i}t \hat{W}),
	\label{eq:qaoa_mixer}
\end{equation}
which is defined by the mixing operator
\begin{equation}
	\hat{W} = \sum_{k,k^\prime=0}^{K-1} w_{kk^\prime} \ket{k}\bra{k^\prime},
	\label{eq:adjacency}
\end{equation}
where typically $w_{kk^\prime} \in \{0, 1\}$. This can be interpreted as implementing a continuous-time quantum walk for time $t \geq 0$ over an undirected graph of $K$ vertices with adjacency matrix $w_{kk^\prime}$, where $w_{kk^\prime} = 1$ if vertices $k$ and $k^\prime$ are connected and $k \neq k^\prime$~\cite{hadfield_quantum_2019,marsh_quantum_2019}. 
For a complete graph $\hat{W}$, one can write
\begin{equation}
  \hat{U}_{W\text{-QAOA}}(t) = e^{it} \left[ \hat{I} + ( e^{-itK} - 1 ) \frac{1}{K} \sum_{k,k'=0}^{K-1} \ket{k} \bra{k'} \right].
\label{eq:qaoa_c_op}
\end{equation}
The action of a single iteration of $\hat{U}_\text{QAOA}(t, \gamma) = \hat{U}_{W\text{-QAOA}}(t)\hat{U}_Q(\gamma)$ then maps the amplitudes of an arbitrary state $\sum_k \alpha_k \ket{k}$ (up to a global phase $e^{it}$) as
\begin{align}
  \alpha_k \hspace{1mm} \mapsto \hspace{1mm} e^{-i \gamma f_k} \alpha_k + ( e^{-i K t} - 1 ) \left( \frac{1}{K} \sum_{k'=0}^{K-1} e^{-i \gamma f_{k'}} \alpha_{k'} \right).
\label{eq:qaoa_c_amplitude}
\end{align}
We see that the second term averages the amplitudes over the entire solution space and is the same for all $k$. Amplification of a particular coefficient $\alpha_k$ then depends on how this local information compares with the global average. This is a useful property in the absence of an identified solution space structure, since $k$ is distinguished solely by the locally phase-encoded $f_k$~\cite{slate_quantum_2021,bennett_quantum_2022}.
Notice that the unbiased coupling in \cref{eq:qaoa_c_op} means that amplitudes at any two points $k$, $k'$ with $f_k \approx f_{k'}$ evolve similarly under $\hat{U}_\text{QAOA}(t, \gamma)$, and will also respond similarly to variation in $t$ and $\gamma$. This is a potential disadvantage in the context of CMOPs since contours in $f$ result in many degenerate $f_k$. Highly degenerate solutions will greatly influence the sum in \cref{eq:qaoa_c_amplitude}, and thus are likely to dominate the optimisation process.

The QAOA was originally defined with the $\hat{W}$ structured according to a hypercube graph, as a hypercube on $M$ qubits is easily implemented as $\sum_ {i=0}^{M-1}\hat{X}^{(i)}$, where superscript $(i)$ denotes action on qubit $i$~\cite{farhi_quantum_2014}. For a hypercube graph $\hat{W}$, the QAOA mixing unitary can be written as:
\begin{align}
  \hat{U}_{W\text{-QAOA}}(t) \nonumber = \sum_{w=0}^M (\cos t)^{M-w} (-i \sin t)^w \sum_{k=0}^{K-1} \sum_{b \in \mathcal{B}_w} \ket{k} \bra{k \oplus b},
\label{eq:qaoa_h_op}
\end{align}
where $\mathcal{B}_w$ is the set of bit strings of Hamming weight $w$, and $k \oplus b$ denotes bitwise XOR between the binary representation of $k$ and $b$. As opposed to \cref{eq:qaoa_c_op}, the hypercube mixer couples points differently according to their respective Hamming distance. Thus, even if there are many points with similar $f_k$ values, amplitudes at such points should only respond similarly to variations in $t$ and $\gamma$ when averages of phase-shifted amplitudes at a fixed Hamming distance away are the same. Given a hypercube embedding of the solution space grid, this is likely to occur primarily when $f$ has particular structural properties, such as rotational symmetry or periodicity.

\begin{figure*}[t]
	\centering
\begin{tikzpicture}
\node[inner sep=0pt] (QMOA) at (0,0)
    {\includegraphics[width=0.32\textwidth]{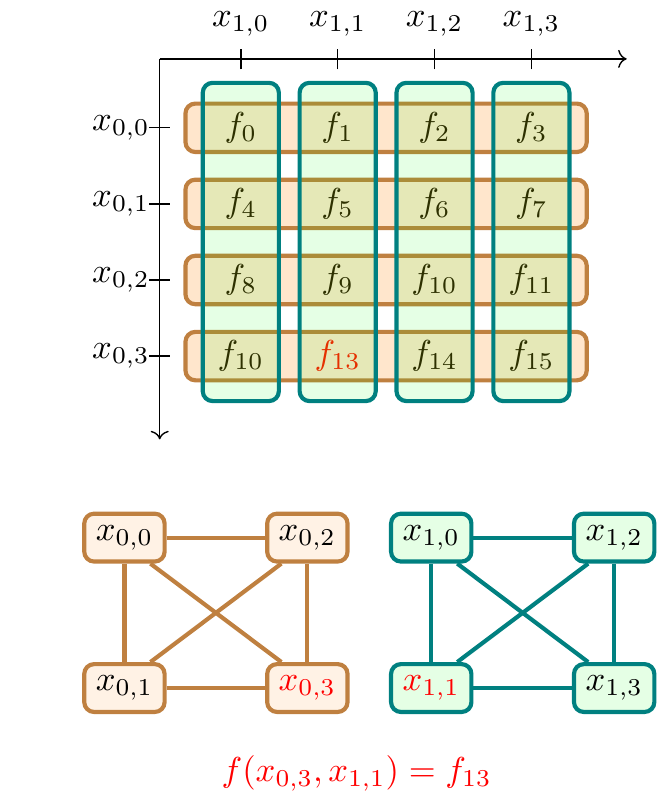}};
\node[inner sep=0pt] (Hypercube_2D) at (4.8,0.2)
    {\includegraphics[width=0.3\textwidth]{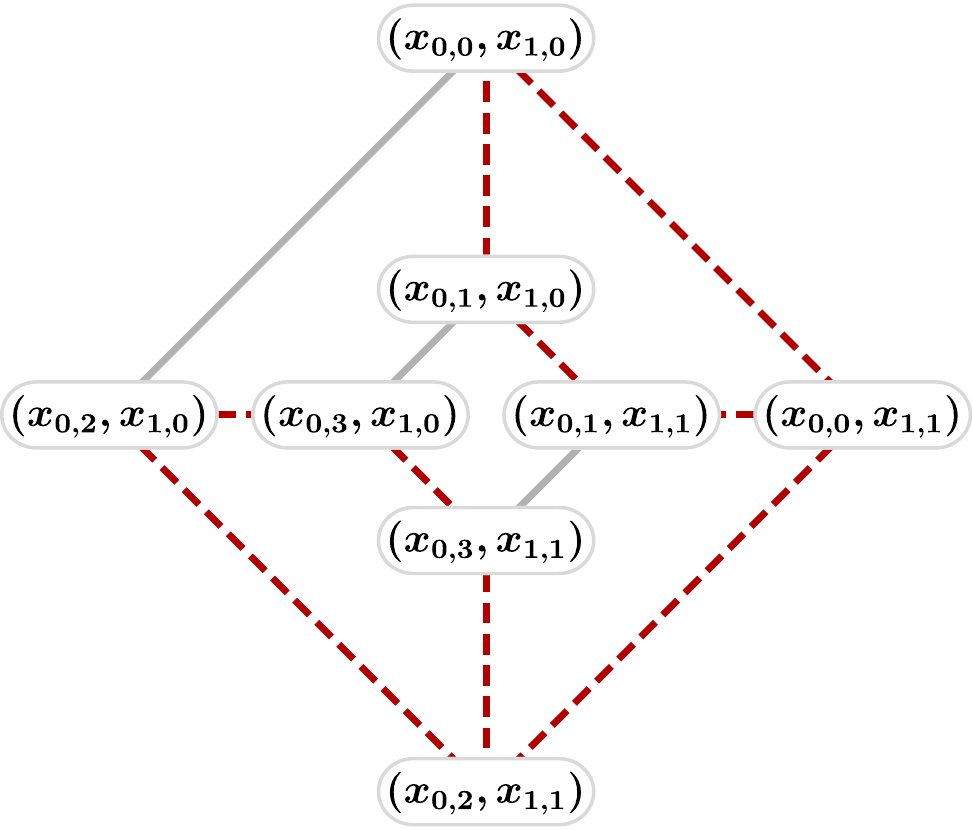}};
\node[inner sep=0pt] (Hypercube_3D) at (10.5,0.2)
    {\includegraphics[width=0.38\textwidth]{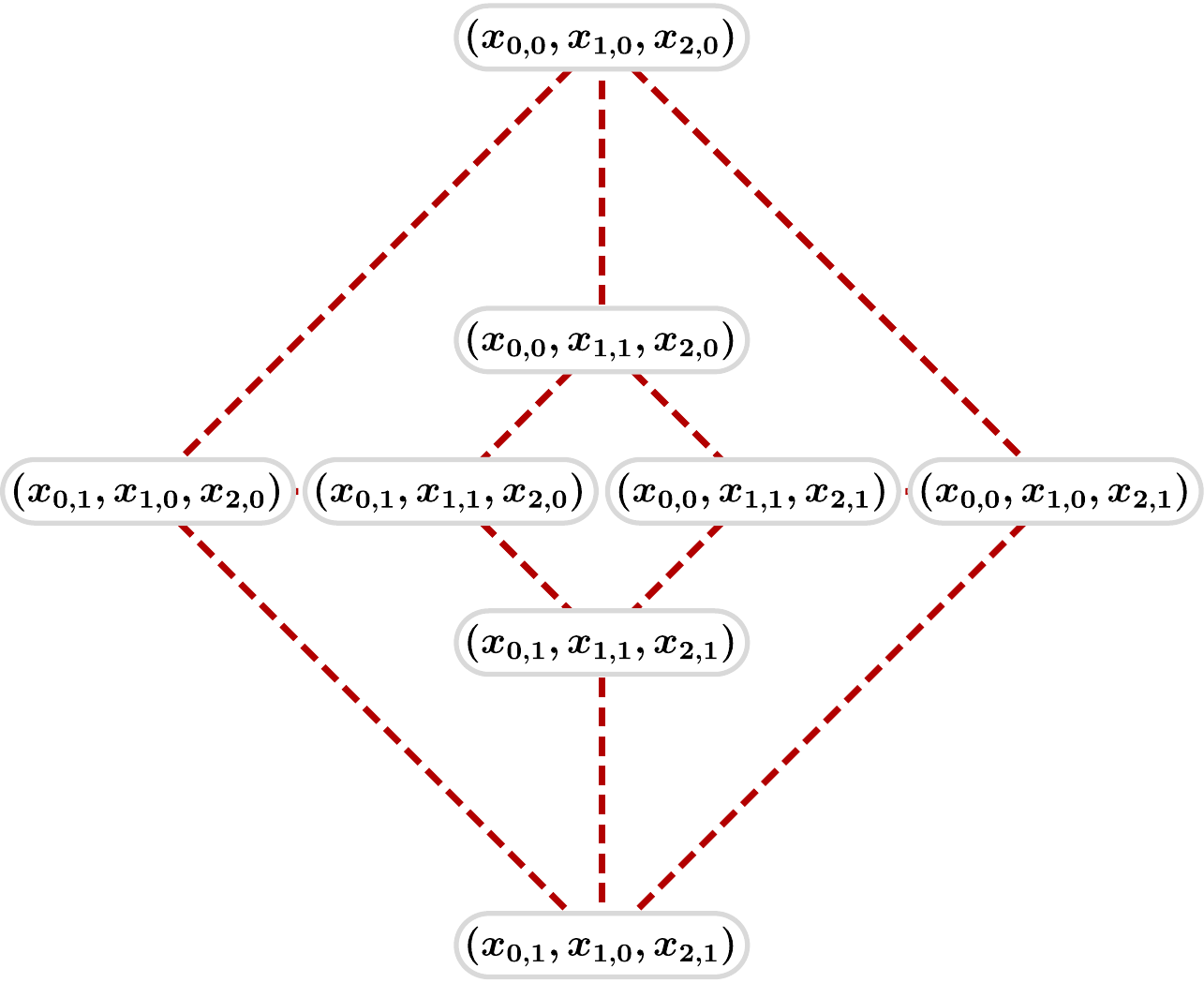}};
    
    \node[] (a) at (0, -3.6)  {{\footnotesize(a)}};
    \node[] (b) at (4.8, -3.6)  {{\footnotesize(b)}};
    \node[] (c) at (10.5, -3.6)  {{\footnotesize(c)}};
\end{tikzpicture}
\caption{(a) Overview of the QMOA for an arbitrary $f$ in $D=2$ discretised over a grid of $16$ points. The horizontal and vertical outlines denote the hyperplanes of constant coordinates. The bottom two graphs in (a) illustrate the coupling between these hyperplanes in $\hat{U}_{W\text{-QMOA}}$ with a complete graph in each dimension. (b,c) Examples of the coupling produced on $\bm{x}_k$ by $\hat{U}_{W\text{-QAOA}}$ with a hypercube $\hat{W}$ are shown for a solution space of size $K=8$ in $D=2$ and $D=3$, respectively. The dashed red line indicates the grid embedding, which in this case is approximate for $D=2$ and exact for $D=3$. For higher-dimensional grids there are classical algorithms that map adjacent grid points to hypercube with maximum distance $\mathcal{O}(M)$~\cite{chan_embedding_1991}. Such algorithms might be implemented as a re-indexing of $\bm{x}_k$, however, we constrain this study to $\bm{x}_k$ as defined for $\ket{k}$.}
\label{fig:mixing_structure}
\end{figure*}
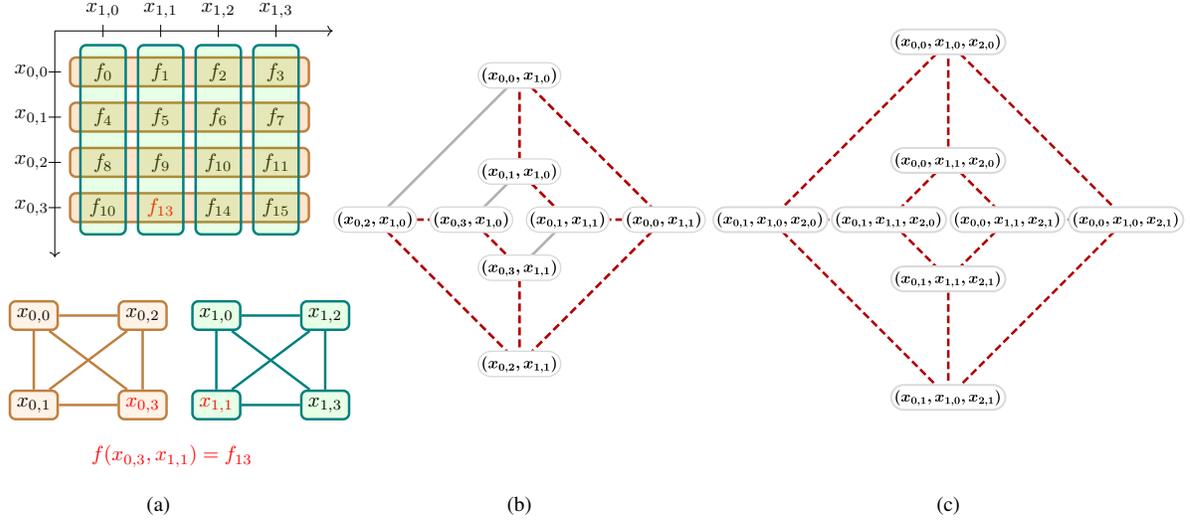

In the context of a quantum search over the discretised solution space of a CMOP, the hypercube has the desirable property of (at least approximate) preservation of the solution space structure, as grids in one, two, and three dimensions can be embedded in a hypercube~\cite{ostrouchov_parallel_1987}. Examples of the grid embedding induced by $\hat{U}_{W\text{-QAOA}}$ are shown in \cref{fig:mixing_structure} (b) and (c). Also, a hypercube graph has a diameter of $M$ and $M$ disjoint paths between any two vertices~\cite{ostrouchov_parallel_1987}, so the distance between any two $\bm{x}_k$ is exponentially smaller than $K$. 

\subsection{Quantum Optimisation with Wavepacket Evolution.} The approach to continuous-variable optimisation described in \cite{verdon_quantum_2019} (also \cite[Sec. III.B]{verdon_universal_nodate}) using continuous quantum variables consists of the propagation of an initial Gaussian wavepacket under a phase-shift followed by the mixing unitary
\begin{equation}
	\hat{U}(t) = \prod_{d=0}^{D-1} e^{-i t \hat{p}_d^2},
	\label{eq:cts_mixer}
\end{equation}
where $\hat{p}_d$ is the momentum operator conjugate to the continuous coordinate $\hat{x}_d$. This choice is inspired by considering the quantum simulation of a particle evolving under the potential $f(\bm{x})$.

Here we examine a discretised form of this algorithm, with the problem solution space encoded in $\ket{k}$. The state is initialised to a discretised Gaussian wavepacket,
\begin{equation}
	\ket{\psi_0}=\frac{1}{\sqrt{A}}\sum_{k=0}^{K-1}\prod_{d=0}^{D-1} e^{-\frac{(x_{k}^{(d)} - \mu_d)^2}{2{\sigma_d}^2}}\ket{k}
	\label{eq:wave-packet}
\end{equation}
where $x_{k}^{(d)}$ is the $d^{th}$ component of $\bm{x}_k$, $\mu_d$ and $\sigma_d$ are the centre and width of the wavepacket in each dimension, and $A$ is a normalising constant. Discretising the mixing unitary requires a discrete form of the continuous momentum operator. For our implementation of QOWE, we construct a discrete analogue of the continuous-variable relationship $\hat{p}_d = \mathcal{F}^{-1} \hat{x}_d \mathcal{F}$ (in each dimension), where $\mathcal{F}$ is the continuous Fourier transform. The continuous Fourier transform along a single dimension can be approximated on the discretised grid as
\begin{equation}
	\mathcal{F} \approx F_d := \sum_{n_d = 0}^{N-1} e^{-i x_{d,0} \kappa_{d,n_d}} \ket{n_d} \bra{n_d} \text{DFT},
\end{equation}
where $\text{DFT}$ is the centred discrete Fourier transform, and $\kappa_{d,n_d} = \kappa_{d,0} + n_d \Delta \kappa_d$ is a momentum-space grid point, with $\Delta \kappa_d = \frac{2\pi}{N \Delta x_d}$, $\kappa_{d,0} = \Delta \kappa_d ( -N + 1 + \lfloor \frac{N-1}{2} \rfloor )$, and $n_d = 0, \dots, N-1$. The corresponding Fourier transform over the entire discretised solution space is then $F := \otimes_{d=0}^{D-1} F_d$, and the mixing unitary is
\begin{equation}
	\hat{U}_{\vert\kappa_{k}\vert^2}(t) = F^{-1}e^{-\text{i}t \hat{W}_{\kappa}} F
	\label{eq:gaussian_mixer}
\end{equation}
where $\hat{W}_{\kappa}$ is the diagonal operator,
\begin{equation}
	\hat{W}_{\kappa}=\sum_{k = 0}^{K-1} \vert\bm{\kappa}_{k}\vert^2 \ket{k}\bra{k},
	\label{eq:momentum_mixer}
\end{equation}
and where $\bm{\kappa}_k = (\kappa_{0,n_0}, \dots, \kappa_{D-1,n_{D-1}})$ is a momentum space grid point with a similar indexing to $\bm{x}_k$.

Applying the phase-shift unitary followed by the first Fourier transform in \cref{eq:gaussian_mixer} and computational basis measurement is related to Jordan's algorithm for gradient computation~\cite{jordan_fast_2005}. Here, the gradient information is used coherently by following the first Fourier transform by the remaining two unitaries in~\cref{eq:gaussian_mixer}, instead of performing a measurement.

\subsection{The Quantum Multivariable Optimisation Algorithm.} The QMOA mixer is taken to be a unitary of separable CTQWs,
\begin{equation}
	\hat{U}_{W\text{-QMOA}}(\bm{t}) = \prod_{d=0}^{D-1}\exp(-\text{i}t_d \hat{C}_d),
	\label{eq:nd_walk}
\end{equation}
where $\bm{t} = (t_0,...t_{D-1})$ with $t_d \geq 0$ and $\hat{C}_d$ is the adjacency matrix of an undirected graph (see \cref{eq:adjacency}) connecting vertices along the dimension $d$.
The discretisation of the QOWE mixer is of a similar form if the generator of \cref{eq:gaussian_mixer} is interpreted as a composite of complete graphs with complex-valued $w_{kk'}$. In QMOA, we only consider cases where $w_{kk'} \in \{ 0, 1 \}$. With $\hat{C}_d$ as a cycle graph, $\hat{W}$ is equivalent to a finite difference approximation of the Laplacian (i.e., a different discretisation of \cref{eq:cts_mixer}). However, we consider more general graphs that do not correspond to different discretisations of \cref{eq:cts_mixer}, but do separate into independent quantum walks in each dimension. The case where each $\hat{C}_d$ is a complete graph is depicted in \cref{fig:mixing_structure} (a) for a two-dimensional $4 \times 4$ grid. For the QMOA, we found that a complete graph $\hat{C}$ achieves the lowest mean error (see App. \ref{app:qmoa_simulation}). 

Under the condition that $\hat{C}_d$ is circulant, and therefore diagonalised by the coefficient matrix of the discrete Fourier transform, \cref{eq:nd_walk} is efficiently realisable as 
\begin{equation}
	(\text{DFT}^{-1})^{\otimes D} \exp\left(- \rm{i} \hat{\Lambda}(\bm{t}) \right) \text{DFT}^{\otimes D},
\end{equation}
where $\text{DFT}$ denotes the discrete Fourier transform and,
\begin{equation}
	\hat{\Lambda}(\bm{t}) = \sum_{d = 0}^{D-1}t_{d}\sum_{n_d=0}^{N - 1}\Lambda_{d,n_d}\ket{x_{d,n_d}}\bra{x_{d,n_d}},
	\label{eq:circulant_mixer}
\end{equation}
is constructed using the closed form solution for the eigenvalues $\Lambda_{d,n_d}$ of $\hat{C}_d$. We note that of the graphs introduced for the QAOA, the complete graph is circulant, while the hypercube graph is non-circulant.  Altogether, the QMOA ansatz unitary $\hat{U}_\text{QMOA}$ has a $\mathcal{O}\left(\text{polylog} \, K\right)$ gate complexity resulting from $D$ instances of the quantum Fourier transform \cite{hales_improved_2000}.

For complete graphs $\hat{C}_d$, the mixing unitary can be written as,
\begin{align}
  \hat{U}_{W\text{-QMOA}}(\bm{t}) \nonumber 
  = \bigotimes_{d=0}^{D-1} e^{i t_d} \left[ \hat{I} + ( e^{-i t_d N} - 1 ) \frac{1}{N} \sum_{n_d, n_d' = 0}^{N-1} \ket{n_d} \bra{n_d'} \right].
\end{align}
In each dimension, the operator $\exp(-i t \hat{C}_d)$ applies an unbiased coupling between all points within each line parallel to coordinate axis $d$. The amplitude of a point then evolves according to the average amplitude along the corresponding line, analogous to \cref{eq:qaoa_c_amplitude}. Combining the walks in each dimension, along with the phase-shift, 
$\hat{U}_\text{QMOA}(t, \gamma)=\hat{U}_{W\text{-QMOA}}(t)\hat{U}_Q(\gamma)$ causes the amplitude of a point to evolve according to the locally phase-encoded $f_k$ relative to averages of phase-shifted amplitudes in the various subspaces containing the point. For example, in \cref{fig:mixing_structure} (a) the coordinate subspaces of $\ket{k=13}$ are the row of points containing $x_{0,3}$ and the column of points containing $x_{1,1}$. By averaging phase-shifted amplitudes among these subspaces, rather than simply over the entire solution space as in \cref{eq:qaoa_c_op}, as well as using different walk times $t_d$ in each dimension, $\hat{U}_{W\text{-QWOA}}$ can break degeneracies resulting from contours in $f$ that are non-parallel to the coordinate axes. More generally, the evolution of amplitudes at any two $\ket{k}$ will respond similarly to variations in $t_d$ and $\gamma$ only if there is similarity in their locally encoded $f_k$ and the averages of the phase-encoded $f_k$ in their respective subspaces, which is likely to occur only when there is a high degree of symmetry in $f$. Furthermore, as the minima of a continuous $f$ are stationary points, every line passing through (or near) a minimum will contain multiple $\bm{x}_k$ with $f_k$ close to the minimum value (provided the discretisation is sufficiently dense). Consequently, the separable CTQWs have the potential to mutually re-enforce convergence to subspaces that contain multiple high-quality solutions.

\begin{figure}[!t]
		
	\centering
    \begin{tikzpicture}
        \node[inner sep=0pt] (A) at (0,0.35)
    {\includegraphics[width=0.15\linewidth]{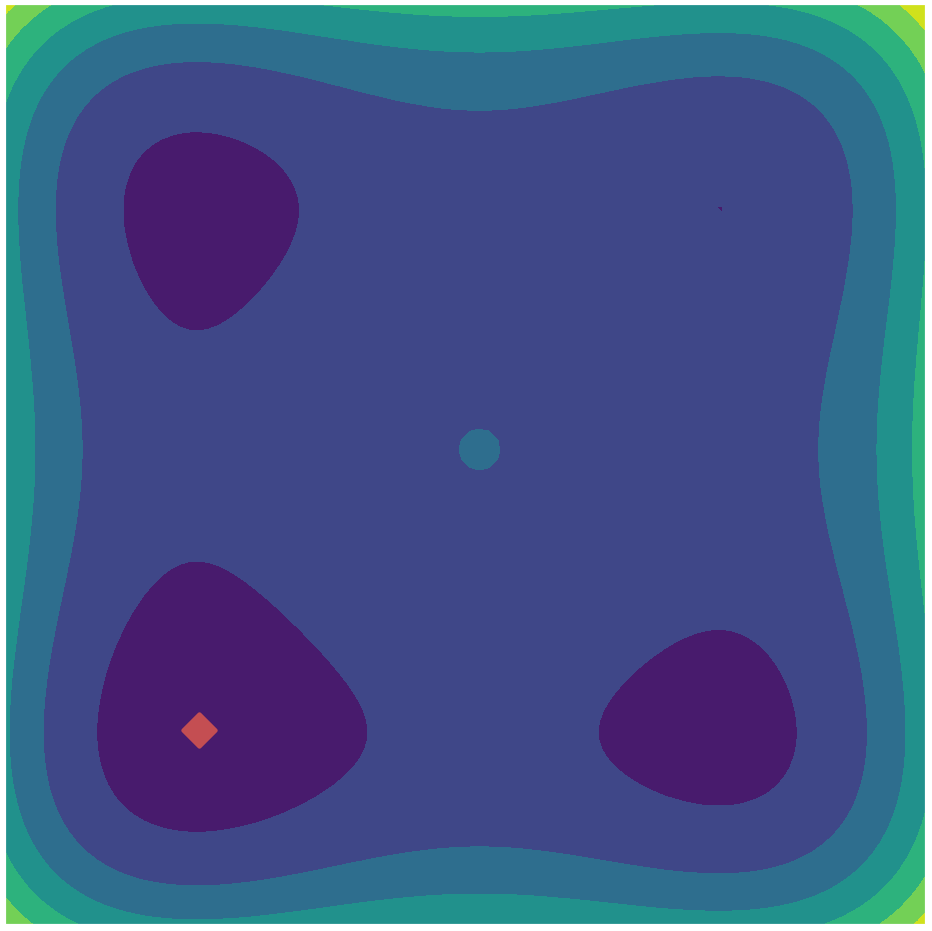}};
    \node[inner sep=0pt] (B) at (3.9,0)
    {\includegraphics[width=0.33\linewidth]{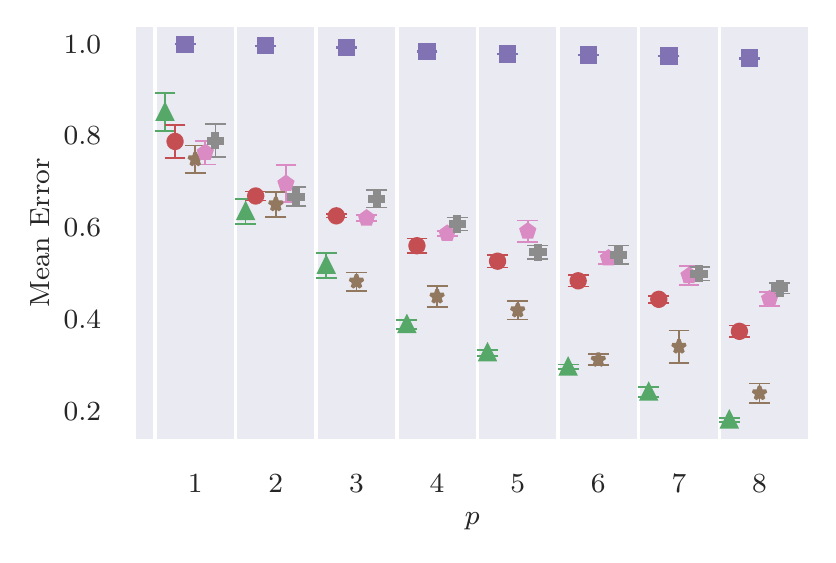}};
    \node[inner sep=0pt] (C) at (8.2,0.35)
    {\includegraphics[width=0.15\linewidth]{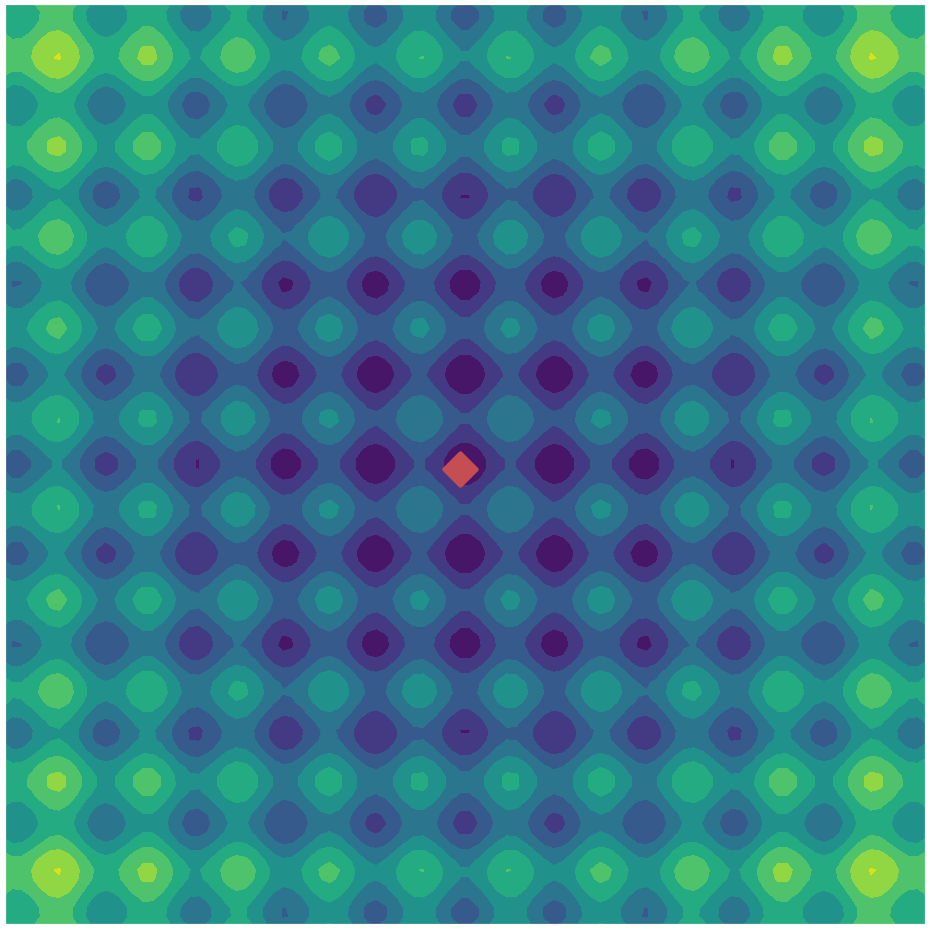}};
    \node[inner sep=0pt] (D) at (12.3,0)
    {\includegraphics[width=0.35\linewidth]{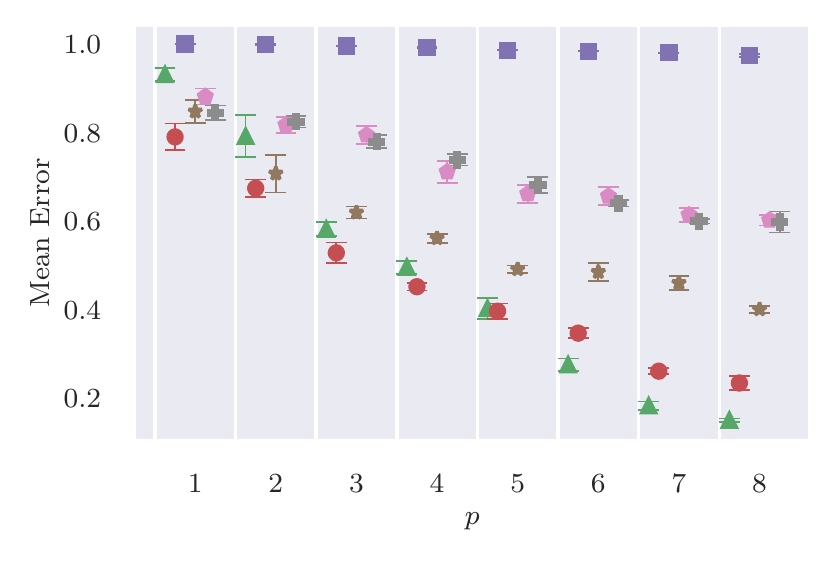}};
    
    \node[inner sep=0pt] (a) at (1.95,-2)
    {{\footnotesize (a) Styblinski-Tang function}};
    \node[inner sep=0pt] (b) at (10.25,-2)
    {{\footnotesize (b) Rastrigin function}};
    
    \node[inner sep=0pt] (L) at (7.0,-2.8)
    {\includegraphics[width=0.55\linewidth]{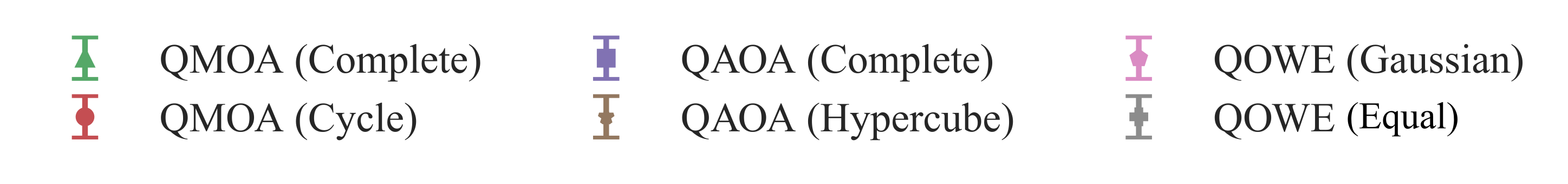}};

    \end{tikzpicture}		

    \vspace{-0.4cm}

	\caption{Contour plots of the (a) STF and (b) RF in $D=2$ (see App.~\ref{app:test_functions}) next to the mean error of QVA minimisation of the two functions in $D=3$ with $N=32$ ($15$ qubits). In each contour plot, a red diamond marks the global minimum.}
	\label{fig:continuous_depth_plots}
		
\end{figure}

\subsection{Comparison of the QMOA, QOWE, and QAOA.} The three QVAs were applied to optimisation of the Styblinski-Tang function (STF), 
\[ f(\boldsymbol{x}) = \frac{\sum_{d=0}^{D-1} x_{d}^{4} - 16x_{d}^{2} + 5x_{d}}{2},\]
and Rastrigin function (RF), 
\[f(\mathbf{x}) = 10 D + \sum_{d=0}^{D-1} \left[x_d^2 - 10 \cos(2 \pi x_d)\right],\]
in $D=3$. These functions were chosen as they are qualitatively different in the number of local minima, rotational symmetry, and $f_k$ magnitude (see \cref{fig:continuous_depth_plots} and App. \ref{app:test_functions}). The specific QVA configurations considered include the QMOA with complete graph $\hat{C}$ (QMOA (Complete)) and cycle graph $\hat{C}$ (QMOA (Cycle)), QOWE with a Gaussian $\ket{\psi_0}$  (QOWE (Gaussian)) and $\ket{\psi_0}$ as an equal superposition (QOWE (Equal)), and the QAOA with a complete graph (QAOA (Complete)) and hypercube (QAOA (Hypercube)) $\hat{W}$. In \cref{fig:continuous_depth_plots}, QMOA (Complete) achieves the lowest mean error for both functions, followed by the QAOA (Hypercube) for the STF and the QMOA (Cycle) for the RF. The mean error is similar for QOWE (Gaussian) and QOWE (Equal). 

\begin{figure*}[th]
	\centering
		
	\subfigure[Styblinski-Tang function]{%
		\includegraphics[width=0.19\linewidth]{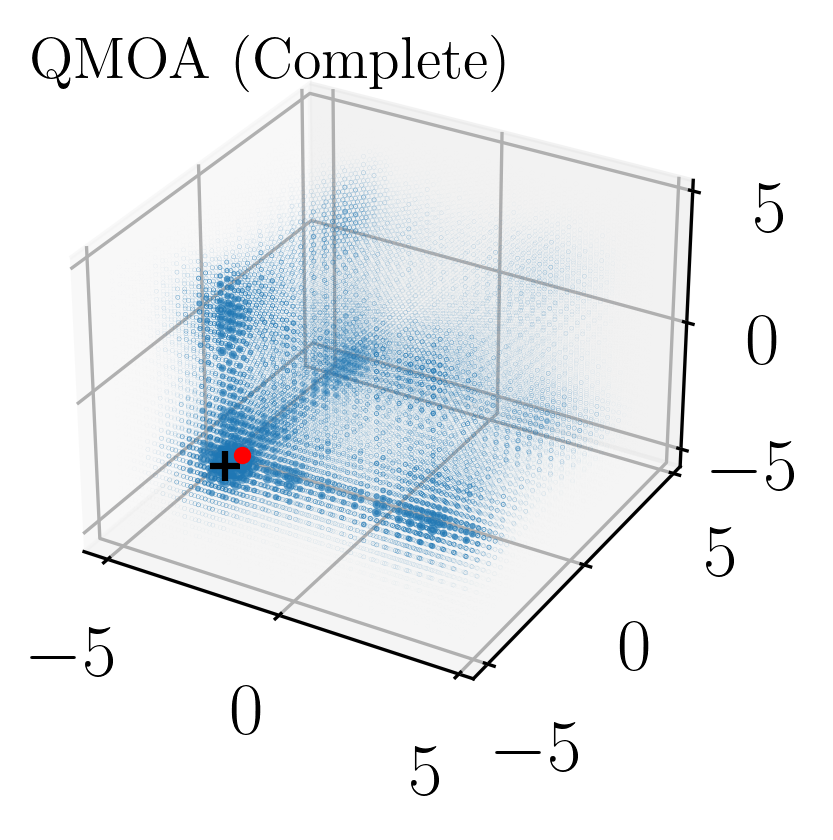}
		\includegraphics[width=0.19\linewidth]{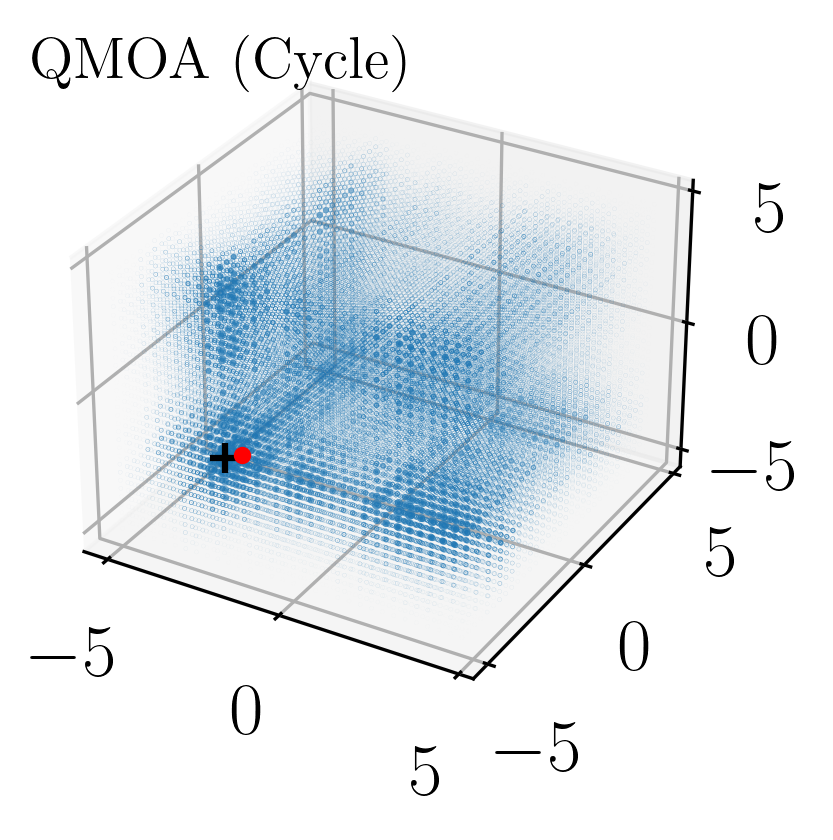}
		\includegraphics[width=0.19\linewidth]{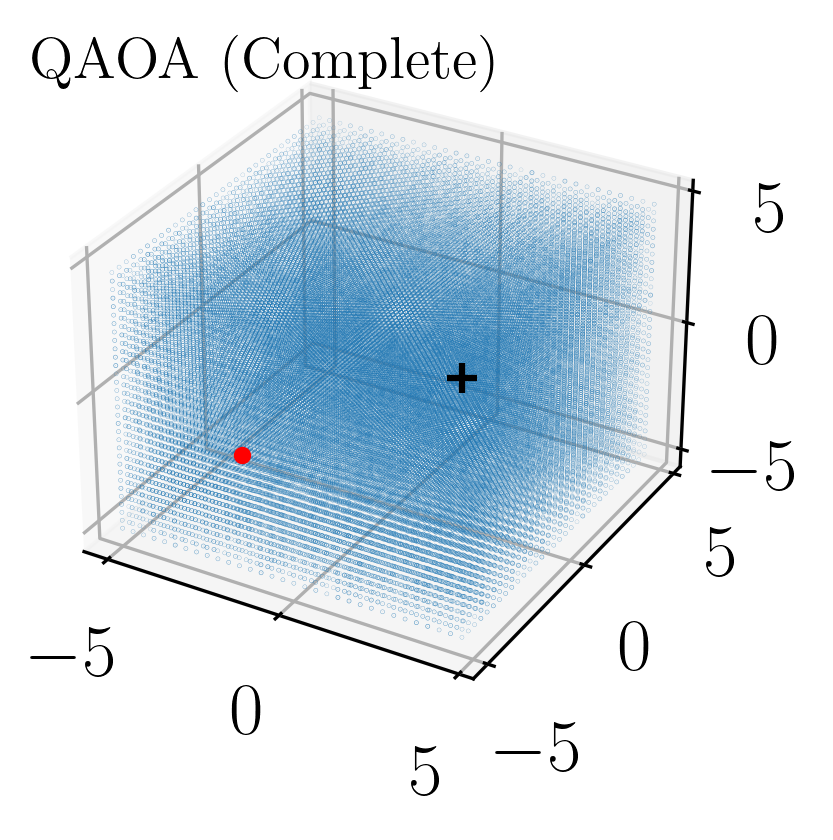}
		\includegraphics[width=0.19\linewidth]{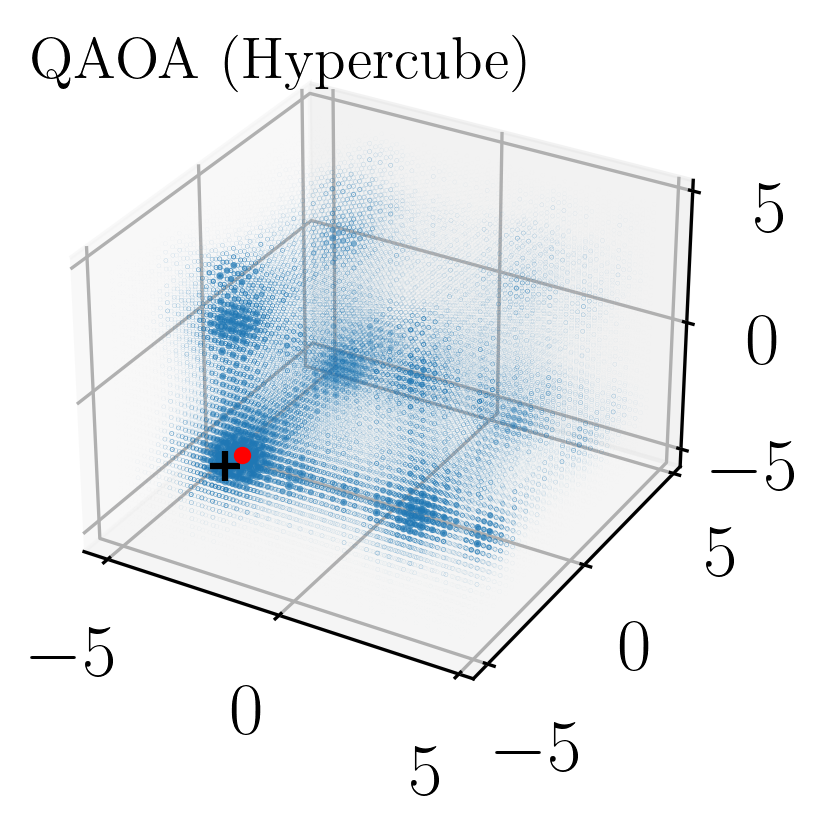}
		\includegraphics[width=0.19\linewidth]{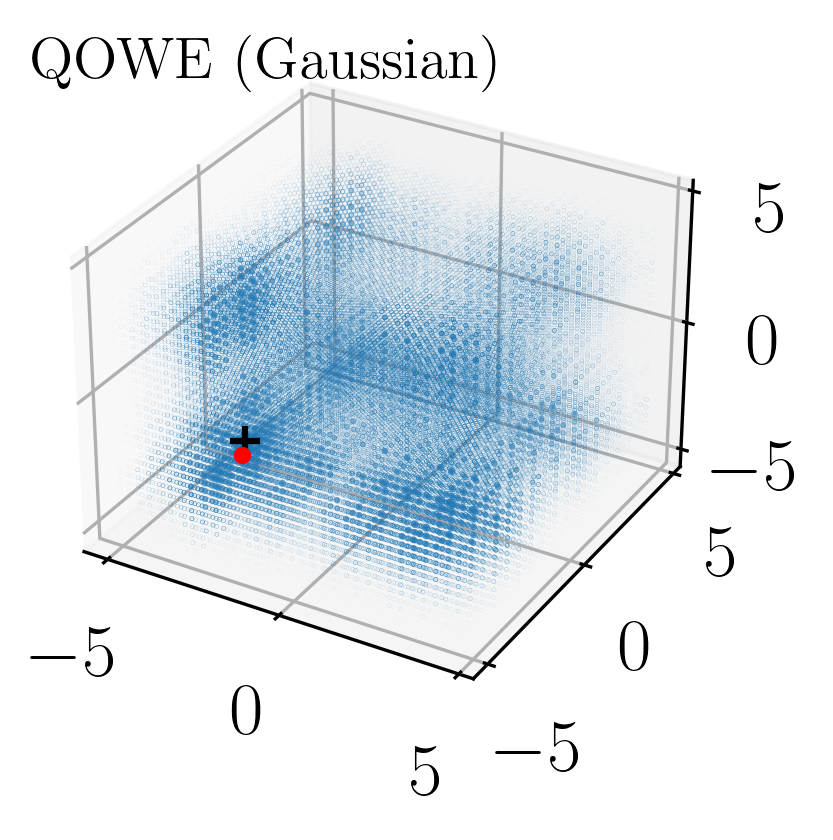}
	}%
		
	\vfill
		
	\subfigure[Rastrigin function]{%
		\includegraphics[width=0.19\linewidth]{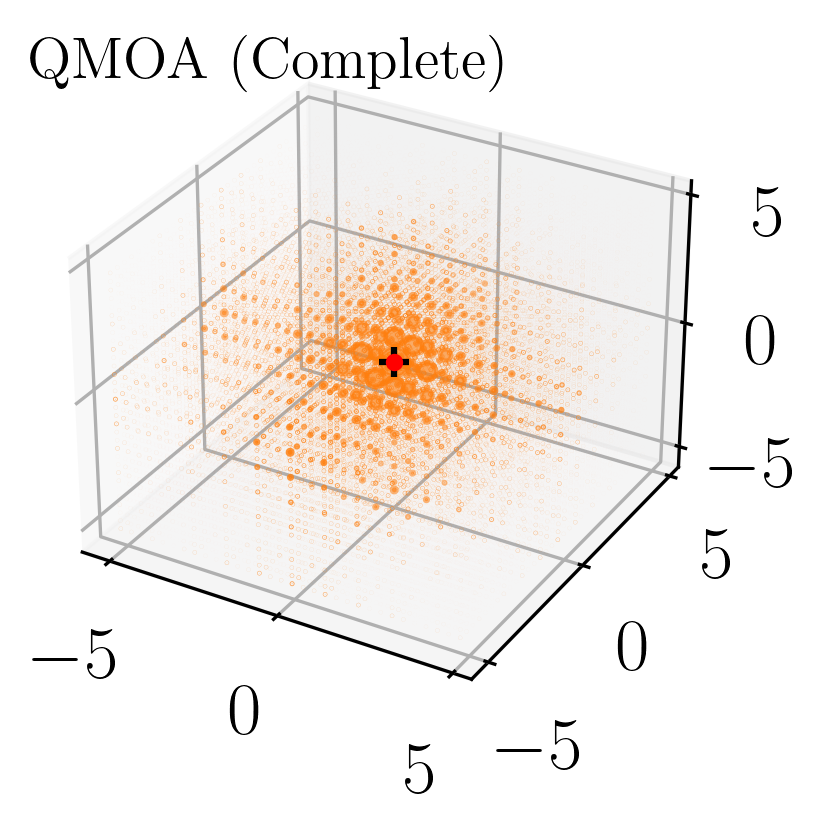}
		\includegraphics[width=0.19\linewidth]{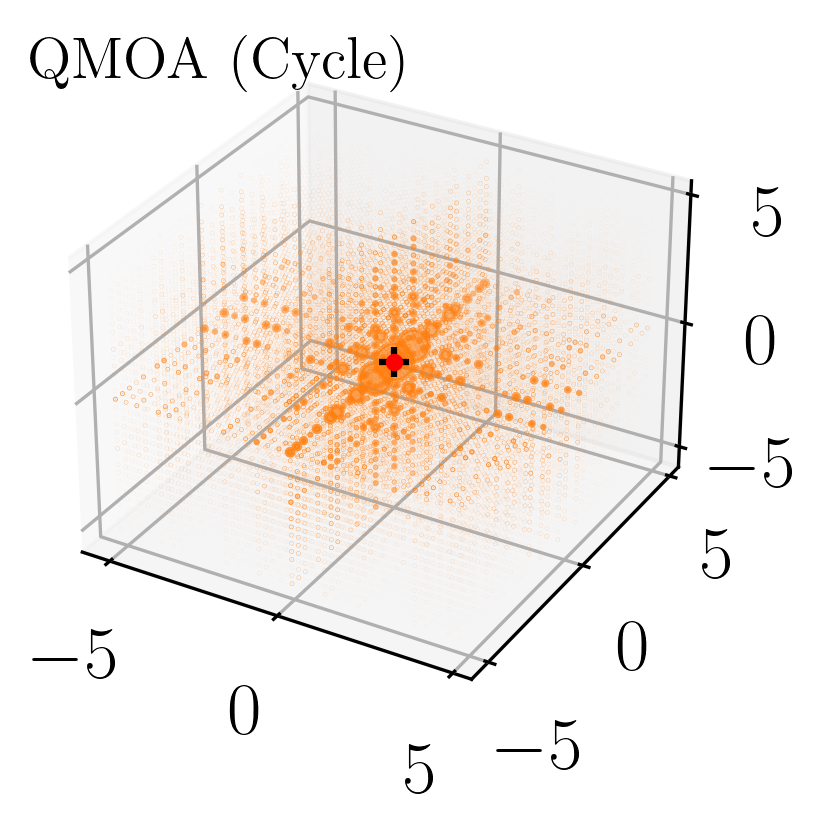}
		\includegraphics[width=0.19\linewidth]{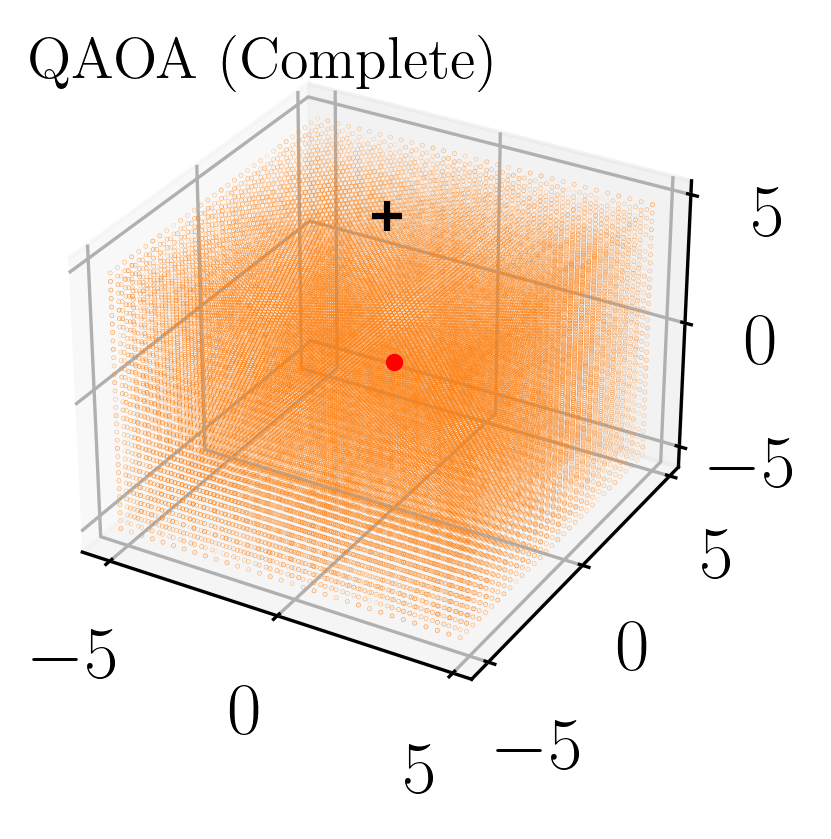}
		\includegraphics[width=0.19\linewidth]{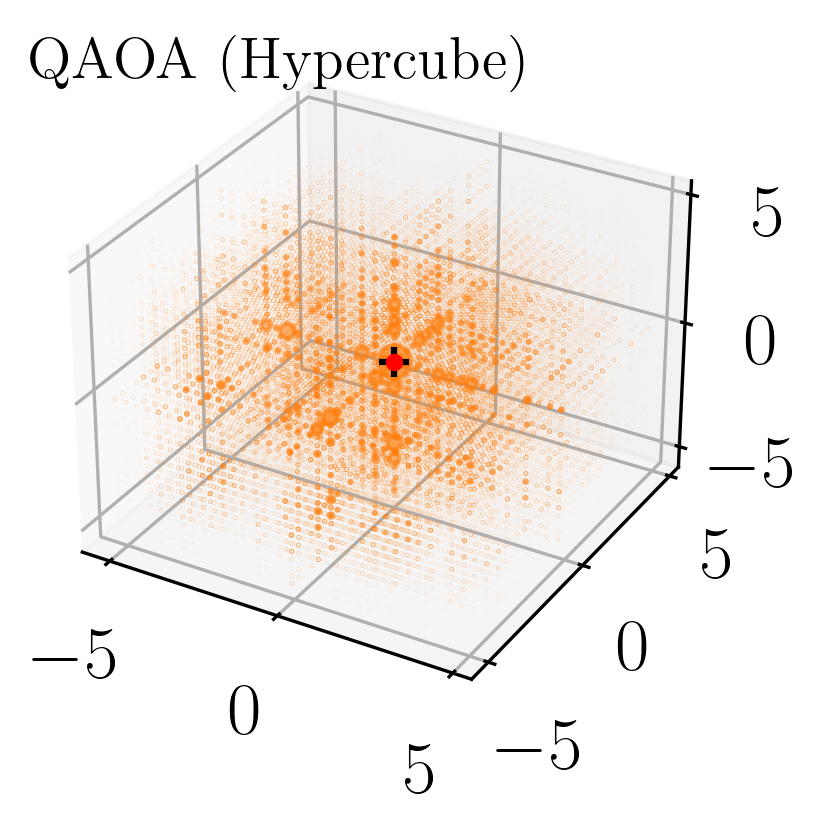}
		\includegraphics[width=0.19\linewidth]{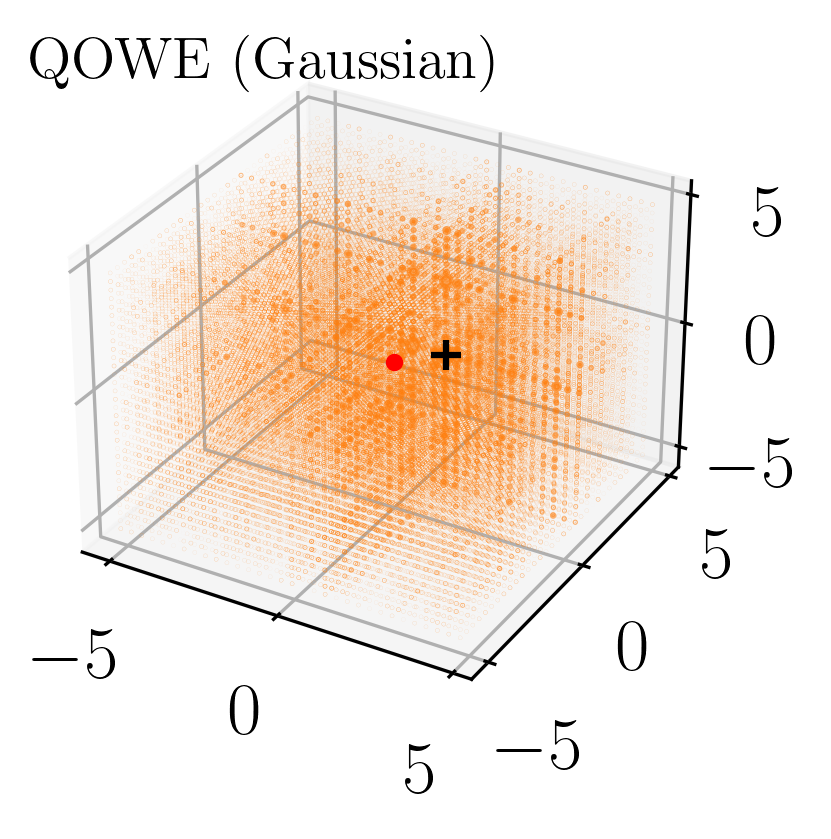}
	}%
	
	\caption{Optimised QVA distributions that achieved the lowest mean error at $p=8$. (a) For the STF, the mean errors are $0.0427$ (QMOA (Complete)), $0.0872$ (QMOA (Cycle)), $0.241$ (QAOA (Complete)), $0.0446$ (QAOA (Hypercube)) and $0.101$ (QOWE (Gaussian State)). (b) For the RF, the mean errors are $0.0657$ (QMOA (Complete)), $0.0889$ (QMOA (Cycle)), $0.447$ (QAOA (Complete)), $0.175$ (QAOA (Hypercube)) and $0.266$ (QOWE (Gaussian State)). QOWE (Equal) is not depicted for either function due to its similarity with QOWE (Gaussian). The red dot indicates the ${\bm{x}^*}$ of the discretised solution space, and the black cross indicates the most amplified $\ket{k}$.}
		
	\label{fig:best_states}
\end{figure*}

The distributions achieving the lowest mean error at $p=8$ are shown in \cref{fig:best_states}. Convergence is strongest for the QMOA (Complete), QMOA (Cycle) and QAOA (Hypercube), with QOWE (Gaussian) appearing diffused in comparison. Consistent with \cref{fig:continuous_depth_plots}, QAOA (Complete) convergence is minimal, consisting of a small transfer of probability density from the initial state. Overall, the QMOA (Complete) states show the most convergence to $\bm{x}_k$ near $\bm{x}^*$. This is most apparent for the RF, where the QMOA (Cycle) converged to $\bm{x}_k$ in a line that contains ${\bm{x}^*}$, and the QAOA (Hypercube) state shows convergence to minima at a greater distance from $\bm{x}^*$. Indeed, for the STF and the RF, the QMOA (Complete) has a statistical distance of $0.132$ and $0.157$, compared to $0.266$ and $0.196$ for QMOA (Cycle) and $0.174$ and $0.264$ for the QAOA (Hypercube).

\begin{figure}[!t]
		
	\centering
		
	\subfigure[Styblinski-Tang function]{%
		\includegraphics[width=0.49\linewidth]{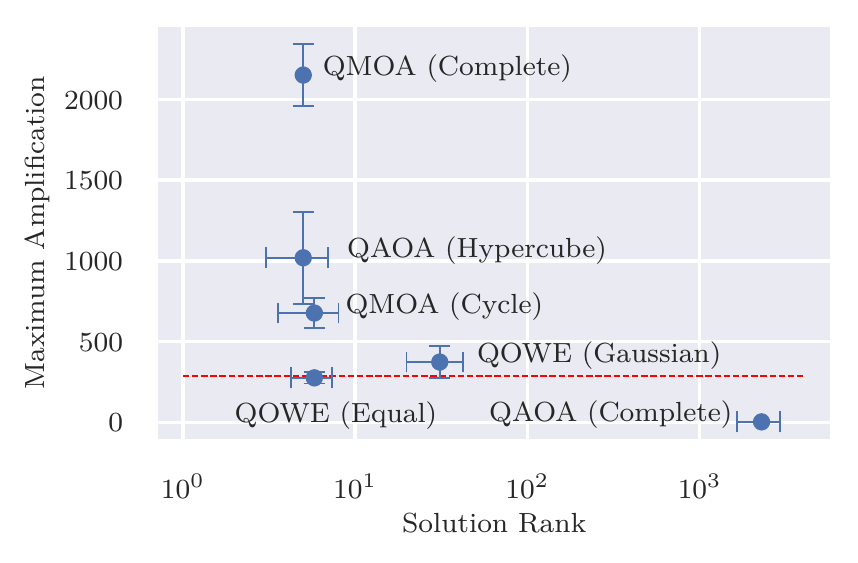}
	}%
	\subfigure[Rastrigin function]{%
		\includegraphics[width=0.49\linewidth]{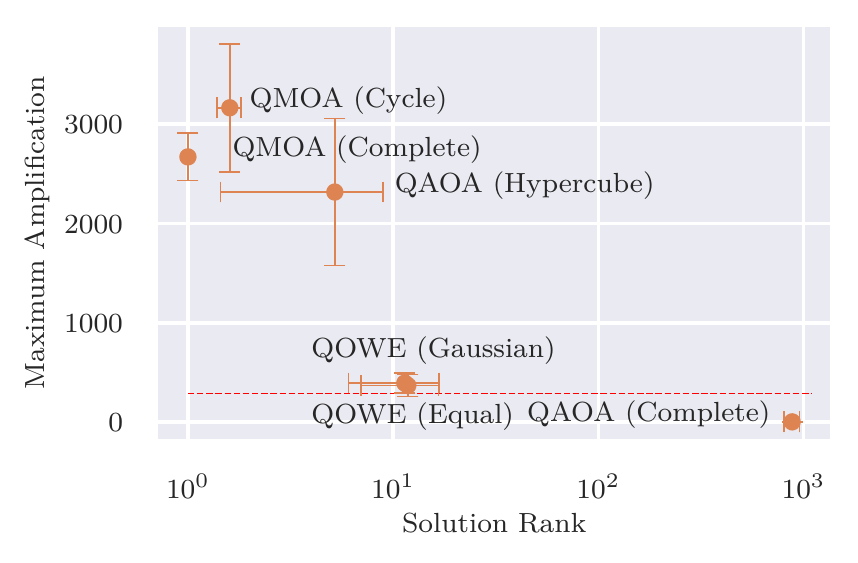}
	}%
	    
	\caption{Mean maximum amplification (see \cref{eq:amplification}) and the corresponding $\bm{x}_k$ rank at $p=8$. A rank of $1$ corresponds to $\bm{x}^*$. The discretised STF had $5887$ unique solutions, and the RF had $1382$ unique solutions. The dashed red line is the state amplification of $288$ achieved by an RDGS at $p=8$.}
		
	\label{fig:amplification}
		
\end{figure}

\cref{fig:amplification} shows that four of the six QVAs have an average state amplification greater than achieved by an RDGS at the same $p$, with the exceptions being the QAOA (Complete) and, for the STF, QOWE (Equal). The QMOA (Complete) and the QMOA (Cycle) have the highest mean amplification for the STF and RF, respectively, with the QMOA (Complete) as a close second for the RF. The QMOA (Complete) converges to the lowest ranked solutions overall (5th lowest for the STF and ${\bm{x}^*}$ for the RF). In contrast, the QMOA (Cycle) and the QAOA (Hypercube) performance varies significantly over the two functions. The QAOA (Complete) performs poorly compared to the other QVAs.

\subsection{Detailed Comparison of the QMOA (Complete) and the QAOA (Hypercube).} 
\begin{figure}[t!]
	\centering
		
	\includegraphics[width=0.495\linewidth]{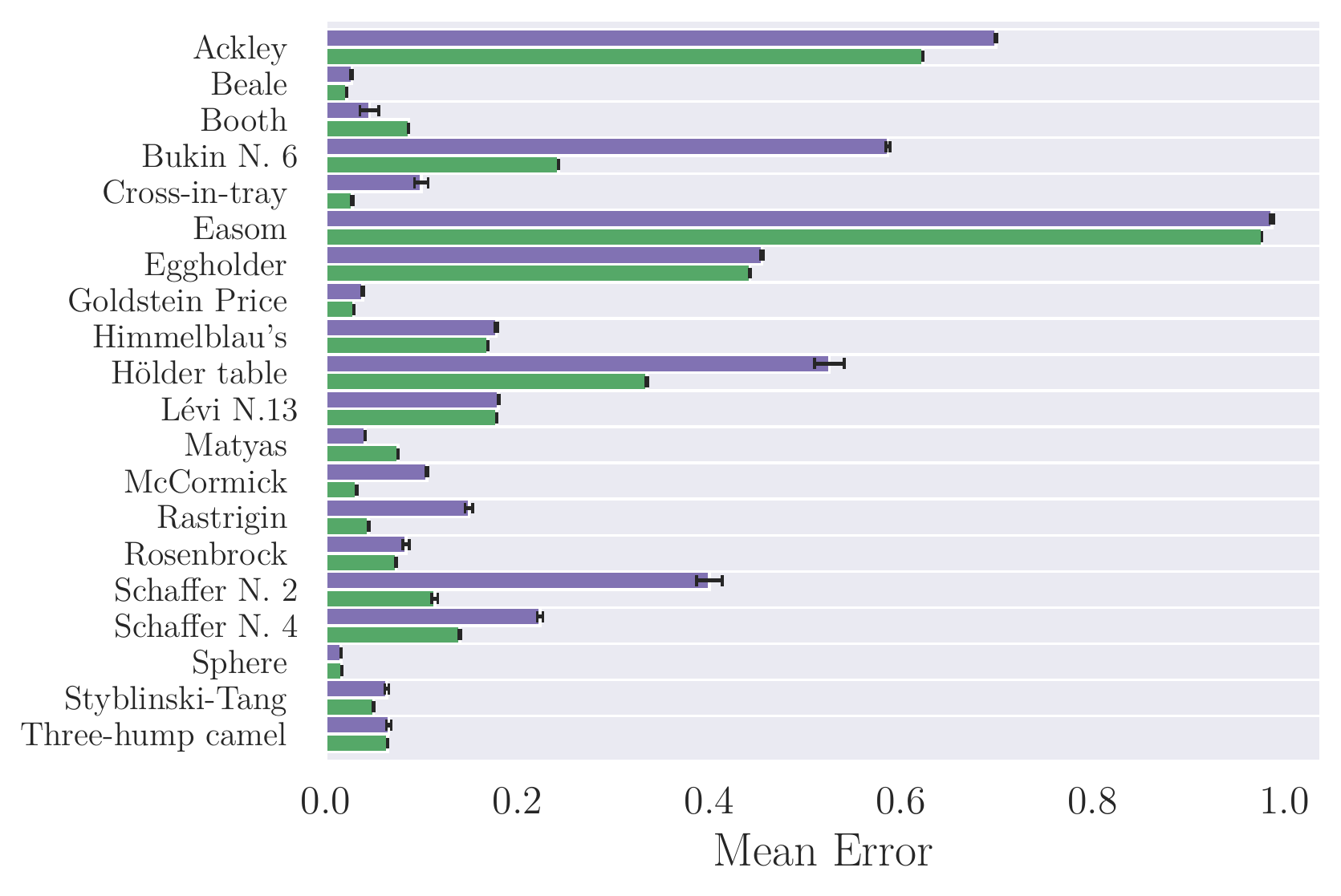}
	\includegraphics[width=0.495\linewidth]{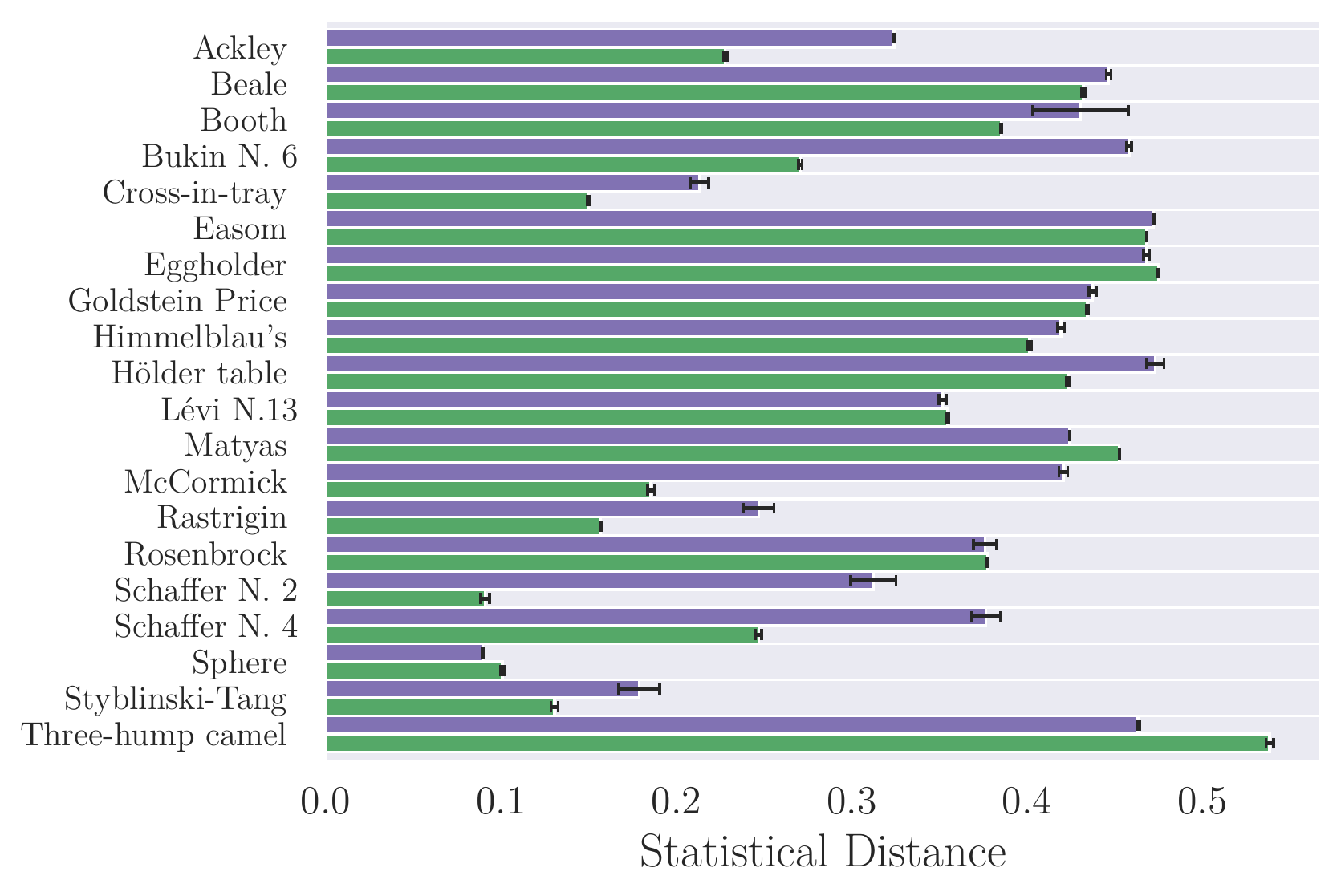}
		
	\includegraphics[width=0.42\linewidth]{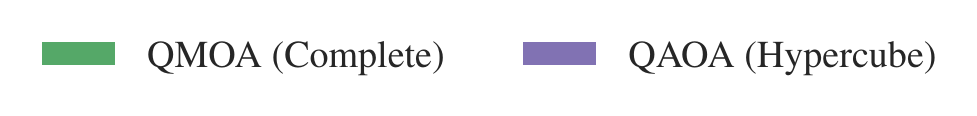}
	
	\caption{(Left) Mean error and (right) statistical distance from ${\bm{x}^*}$ for the QMOA (Complete) and the QAOA (Hypercube) on the $D=2$ test function set with $N=32$ ($15$ qubits) at $p=8$.}
		
	\label{fig:qmoa_vs_qaoa}
		
\end{figure}

\cref{fig:qmoa_vs_qaoa} shows that the QMOA (Complete) achieved the lowest mean error and statistical distance for $16$ of $20$ test-functions at $D=2$. The QAOA (Hypercube) achieved a slightly lower mean error or statistical distance on functions with a single minimum in the search domain (the Matyas, Sphere, Three-Hump Camel and Booth functions). Performance was close (in favour of the QMOA (Complete)) on functions with three or fewer local minima in the search domain (the Easom, Himmelblau's, L\'evi N.13, Goldstein-Price, Beal, Rosenbrock and Eggholder functions). An exception to this pattern is the Eggholder function. The difference between the QMOA (Complete) and the QAOA (Hypercube) is greatest in favour of the QMOA (Complete) for oscillatory functions (the Schaffer N. $2$, Schaffer N. $4$, H\"older table, Cross-in-tray, Ackley and Rastrigin functions), functions with many local minima (the Styblinski-Tang function), or functions with a narrow `valley' across much of the search domain (the McKormic and Bukin functions).

To examine the scaling behaviours of the QMOA (Complete) and the QAOA (Hypercube), we considered the maximum amplification (see \cref{eq:amplification}) achieved in optimisation of the RF with increasing $D$ and $N$, assuming a scaling relationship, 
\begin{equation}
  \text{maximum amplification} = C \, p^{\alpha D}.
  \label{eq:scaling}
\end{equation}  
We are interested in the variation of $\alpha$ with $D$ and $N$ as $\alpha$ describes the QVA scaling with $p$. These are shown in \cref{tab:parameters}. For the QMOA (Complete) and the QAOA (Hypercube) with $N=16$ and $N=32$, the $\alpha$ is constant or increasing with $D$, which indicates maximum amplification that grows $\sim p^D$.  The QMOA (Complete) $\alpha$ is approximately double the QAOA (Hypercube) $\alpha$ at all $N$. In App. \ref{app:sampling} we see that, for the QMOA (Complete), the observed scaling is consistent with efficient optimisation at higher $D$ and low $p$ even when accounting for measurement overhead in optimisation of $(\bm{t}, \bm{\gamma})$. 

\begin{table}[h]
\centering
{\small
\begin{tabular}{cccccccc}
                          &                 & QMOA (Complete) &                 &                 & QAOA (Hypercube) &                 &     \\
\multicolumn{1}{c|}{$N$}  & $2$             & $3$             & $4$             & $2$             & $3$              & $4$             & $D$ \\ \cline{2-8} 
\multicolumn{1}{c|}{$16$} & $1.07 \pm 0.07$ & $1.04 \pm 0.07$ & $1.04 \pm 0.06$ & $0.41 \pm 0.06$ & $ 0.40 \pm 0.06$  & $0.42 \pm 0.07$ &     \\
\multicolumn{1}{c|}{$32$} & $0.90 \pm 0.06$ & $1.32 \pm 0.09$ &                 & $0.43 \pm 0.05$ & $0.7 \pm 0.1$  &                 &     \\
\multicolumn{1}{c|}{$64$} & $1.12 \pm 0.05$ &                 &                 & $0.34 \pm 0.02$ &                  &                 &    
\end{tabular}%
}
\caption{(Left) The QMOA (Complete) and (right) QAOA (Hypercube) fitted values for $\alpha$ in \cref{eq:scaling} for optimisation of the RF. Fitting curves are shown in App. \ref{app:fitting}.}
\label{tab:parameters}
\end{table}

\clearpage

\section{Discussion} \label{sec:conclusion}

\noindent For all $p$ in \cref{fig:continuous_depth_plots,fig:best_states} the QMOA (Complete and Cycle), QOWE and the QAOA (Hypercube) outperformed the unstructured QAOA (Complete) in minimisation of mean error and statistical distance from $\bm{x}^*$. In \cref{fig:amplification} we see that the QMOA (Complete and Cycle) and the QAOA (Hypercube) exceed the theoretical limit for an unstructured quantum search by orders of magnitude. For optimisation of the STF and RF, the QMOA (Complete and Cycle) and QAOA (Hypercube) outperform QOWE significantly.  As QOWE and QMOA (Complete) have a similar walk evolution and coupling structure, it can be reasoned that the complex edge-weights of the QOWE $\hat{W}$ have a diffusing effect as they result in $\hat{U}_{W\text{-QOWE}}$ having an inherently non-uniform action across lines in the solution space.

In \cref{fig:qmoa_vs_qaoa} the QMOA (Complete) and the QAOA (Hypercube) generalise effectively over $20$ qualitatively diverse test-functions. The QMOA (Complete) provides an advantage for oscillatory functions,  functions with a `valley' parallel to the coordinate axes, and functions with multiple local minima. The one exception is the Eggholder function, which has four distinct oscillatory regions. These are angled anti-symmetrically at approximately $45$ degrees to the coordinate axes, which may negatively impact the potential of the QMOA sub-searches to be mutually re-enforcing. However, overall our results demonstrate that the QMOA (Complete) is the most efficient QVA for CMOPs.

For CMOPs over discretised coordinates, the number of solutions grows exponentially with $D$. Thus, an ideal QVA should have a maximum amplification also growing exponentially in $D$. In contrast, an RDGS at low $p$ and large $K$ produces amplification which grows $\sim p^2$~\cite{bennett_quantum_2022}. Consequently, to achieve the same degree of amplification, the $p$ of an RDGS must increase exponentially with $D$. The empirical scaling results in \cref{tab:parameters} demonstrate $\sim p^D$ scaling for the QMOA (Complete) and the QAOA (Hypercube) over the range of considered $D$. While this result is for one function and over a range of $D$ that is limited by simulation constraints, this is encouraging evidence for the ability of the QMOA (Complete) and the QMOA (Hypercube) to mitigate the `curse of dimensionality' in CMOPs. 

All QVAs incur an overhead due to the state preparations required to optimise the variational parameters. However, existing algorithms based on an RDGS depend on marking solutions below a given threshold value~\cite{bennett_quantum_2022}. Doing so may be inappropriate for CMOPs as the resulting search does not distinguish between minima close to or far from ${\bm{x}^*}$. An alternative approach is to identify an optimal marked set via bisection at the expense of additional computational overhead. Considering these factors, the QMOA (Complete) or the QAOA (Hypercube) may be the preferable choice for CMOPs. Through simulation, we have explored the behaviour of QVAs for CMOPs in terms of function dimension and grid size. The chosen test-function set is qualitatively diverse and thus presents a realistic indication of the general advantages and limitations of the QAOA, QOWE and the QMOA in CMOPs.

\section{Methods}


\subsection{QVA Performance Metrics.} In comparing the performance of QVAs, we are chiefly concerned with the relative ability of each algorithm to minimise $\E{Q}$ at equivalent $p$. For comparison across dissimilar $f$, we use the metric
\begin{equation}
	\text{mean error} \coloneqq \frac{\E{Q} - \min\left(f_k\right)}{\max\left(f_k\right) - \min \left(f_k \right)}.
\end{equation}

In addition, a desirable property specific to CMOPs is the concentration of probability about ${\bm{x}^*}$, as convergence to a local minimum near ${\bm{x}^*}$ is preferred over convergence to an equivalently-valued local minimum at a greater distance. To quantify this property, we consider the statistical distance from ${\bm{x}^*}$,
\begin{equation}
	\text{distance} \coloneqq \frac{1}{\text{dist}_{{\bm{x}^*}}\left(\bm{x}_{k^\prime}\right)} \sum_{k=0}^{K-1}\text{dist}_{{\bm{x}^*}}\left(\bm{x}_k \right)\left\vert\E{k \,\vert \,\bm{t}, \bm{\gamma}}\right\vert^2,   
\end{equation}
where $\text{dist}_{{\bm{x}^*}}(\bm{x}_k) = \lVert {\bm{x}^*} - \bm{x}_k \rVert_2$, $\bm{x}_{k^\prime}$ is a point furthest from ${\bm{x}^*}$.

Finally, to test for the efficient exploitation of solution space structure, we compare the maximum QVA state amplification,
\begin{equation}
 \text{maximum amplification} := \text{max}_k \left(\frac{\vert\E{k \, \vert \, \bm{t}, \bm{\gamma}}\vert^2}{1/K} \right),
	\label{eq:amplification}
\end{equation}
where $1/K$ is the probability of measuring $\ket{k}$ in an unbiased sampling of the solution space. Quantum variational algorithm amplification greater than the amplification of a restricted-depth Grover's search at the same $p$ is proof of convergence due to a structured search. Such amplification is efficient if the highest amplified $\ket{k}$ corresponds to an $\bm{x}_k$ near a minimum, with minima closer to ${\bm{x}^*}$ preferred.

\subsection{Simulation.} The QMOA, QOWE and the QAOA were simulated using the QuOp\_MPI software package~\cite{matwiejew_quop_mpi_2022}. To compute the action of $\hat{U}_{\vert\kappa_{k}\vert^2}$ and $\hat{U}_{W\text{-QMOA}}$, QuOp\_MPI was extended to include the $D$-dimensional FFT methods of the FFTW library~\cite{frigo_fastest_1997,frigo_design_2005}. The Nelder-Mead algorithm implementation included with SciPy 1.7.6 was used for optimisation of $(\bm{t}, \bm{\gamma})$~\cite{gao_implementing_2012}. In Figs. \ref{fig:continuous_depth_plots}--\ref{fig:qmoa_vs_qaoa} and \cref{tab:parameters} the number of Nelder-Mead iterations was set to a maximum of $10^6$ and the adaptive parameter scheme of Gao et al. was enabled~\cite{gao_implementing_2012}. Code for all experiments is included with the latest version of QuOp\_MPI \cite{quop_mpi}. 

For each QVA, multiple simulation runs were conducted for each $p$. The $(\bm{t}, \bm{\gamma})$ that achieved the lowest $\E{Q}$ for depth $p$ were used as initial parameter values for the first $p$ ansatz iterations of $\hat{U}(\bm{t}, \bm{\gamma})$ for the simulation of depth $p+1$. The remaining parameters for the QMOA and the QAOA, $(\bm{t}, \bm{\gamma})$, were initialised from the uniform distributions of $[0, 2 \pi)$ for $t$ and $[-2 \pi, 2 \pi)$ for $\gamma$. The presented results are the mean of ten repeats unless otherwise noted, and uncertainty is reported as the population standard deviation when presented. These experiments are intended to present the limiting performance of the QVAs given (presumably) optimal $(\bm{t}, \bm{\gamma})$.
	
For QOWE simulations, the initial $\mu_d$ were generated from a uniform distribution over
\[\left[\min(\bm{x}_d) + \frac{L_d}{8}, \max(\bm{x}_d) - \frac{L_d}{8} \right],\]
where $L_d = \max(\bm{x}_d) - \min(\bm{x}_d) $, so as to avoid initialisation of the wave-packet close to the solution space boundaries. The initial widths were set to $\sigma_d = \frac{1}{\sqrt{2}}$. Unconstrained optimisation of $(\bm{t}, \bm{\gamma})$ was observed to result in highly diffused and non-convergent $\ket{\bm{t},\bm{\gamma}}$. As such, optimisation was initially conducted with $(\bm{t}, \bm{\gamma})$ constrained to $(0, t_0 + b)$ and $(\gamma_0 - b, \gamma_0 + b)$ with $b=0.1$ and initial values $t_0=\gamma_0=0.1$ (the hand-selected parameter values in~\cite{verdon_quantum_2019}). If the optimised $(\bm{t}, \bm{\gamma})$ had $t$ or $\gamma$ equal to the constraint bounds, $b$ was increased by a factor of $1.2$ and the simulation repeated. For $N=32$ and $p=1$ to $8$, $b = 2 \pi$ was found to produce reliable QOWE convergence. Preliminary results found that QOWE performance was improved by introducing dimensionally independent parameterisation of $t$ (per \cref{eq:nd_walk}). As such, all reported QOWE results have $p(D + 1)$ variational parameters. 

\subsection{Analysis of Scaling Behaviour.} The $\alpha$ and $C$ coefficients in \cref{eq:scaling},  were found by curve fitting, \[\log_2 (\text{maximum amplification}) = \log_2 C + \alpha D \log_2 p,\]
over the mean maximum amplification from $p=1$ to $6$ over $60$ repeats. At each $p$ with the best $(\bm{t}, \bm{\gamma})$ out of the previous $60$ repeats used for the initial variational parameters as previously described. The `curve\_fit' method in  SciPy 1.7.6 was used to obtain $\alpha$ and $C$ by a least-squares fit. The reported uncertainty for $\alpha$ is the standard deviation of the coefficient estimate.  

\section{Acknowledgements}

This work was supported by resources provided by the Pawsey Supercomputing Centre with funding from the Australian Government and the Government of Western Australia. EM acknowledges the support of the Australian Government Research Training Program Scholarship. 

\bibliography{bibliography}

\begin{thebibliography}{10}
\expandafter\ifx\csname url\endcsname\relax
  \def\url#1{\texttt{#1}}\fi
\expandafter\ifx\csname urlprefix\endcsname\relax\def\urlprefix{URL }\fi
\providecommand{\bibinfo}[2]{#2}
\providecommand{\eprint}[2][]{\url{#2}}

\bibitem{matthews_how_2021}
\bibinfo{author}{Matthews, D.}
\newblock \bibinfo{title}{How to get started in quantum computing}.
\newblock \emph{\bibinfo{journal}{Nature}} \textbf{\bibinfo{volume}{591}},
  \bibinfo{pages}{166} (\bibinfo{year}{2021}).

\bibitem{bellman_dynamic_1957}
\bibinfo{author}{Bellman, R.}
\newblock \emph{\bibinfo{title}{Dynamic {Programming}}}
  (\bibinfo{publisher}{Princeton University Press}, \bibinfo{year}{1957}).

\bibitem{farhi_quantum_2014}
\bibinfo{author}{Farhi, E.}, \bibinfo{author}{Goldstone, J.} \&
  \bibinfo{author}{Gutmann, S.}
\newblock \bibinfo{title}{A {Quantum} {Approximate} {Optimization}
  {Algorithm}}.
\newblock \bibinfo{howpublished}{Preprint at https://arxiv.org/abs/1411.4028}
  (\bibinfo{year}{2014}).

\bibitem{hadfield_quantum_2019}
\bibinfo{author}{Hadfield, S.} \emph{et~al.}
\newblock \bibinfo{title}{From the {Quantum} {Approximate} {Optimization}
  {Algorithm} to a {Quantum} {Alternating} {Operator} {Ansatz}}.
\newblock \emph{\bibinfo{journal}{Algorithms}} \textbf{\bibinfo{volume}{12}},
  \bibinfo{pages}{34} (\bibinfo{year}{2019}).

\bibitem{marsh_combinatorial_2020}
\bibinfo{author}{Marsh, S.} \& \bibinfo{author}{Wang, J.~B.}
\newblock \bibinfo{title}{Combinatorial optimization via highly efficient
  quantum walks}.
\newblock \emph{\bibinfo{journal}{Physical Review Research}}
  \textbf{\bibinfo{volume}{2}}, \bibinfo{pages}{023302} (\bibinfo{year}{2020}).

\bibitem{guerreschi_practical_2017}
\bibinfo{author}{Guerreschi, G.~G.} \& \bibinfo{author}{Smelyanskiy, M.}
\newblock \bibinfo{title}{Practical optimization for hybrid quantum-classical
  algorithms}.
\newblock \bibinfo{howpublished}{Preprint at https://arxiv.org/abs/1701.01450}
  (\bibinfo{year}{2017}).

\bibitem{marsh_quantum_2019}
\bibinfo{author}{Marsh, S.} \& \bibinfo{author}{Wang, J.~B.}
\newblock \bibinfo{title}{A quantum walk-assisted approximate algorithm for
  bounded {NP} optimisation problems}.
\newblock \emph{\bibinfo{journal}{Quantum Information Processing}}
  \textbf{\bibinfo{volume}{18}}, \bibinfo{pages}{61} (\bibinfo{year}{2019}).

\bibitem{matwiejew_quop_mpi_2022}
\bibinfo{author}{Matwiejew, E.} \& \bibinfo{author}{Wang, J.~B.}
\newblock \bibinfo{title}{{QuOp}\_mpi: {A} framework for parallel simulation of
  quantum variational algorithms}.
\newblock \emph{\bibinfo{journal}{Journal of Computational Science}}
  \textbf{\bibinfo{volume}{62}}, \bibinfo{pages}{101711}
  (\bibinfo{year}{2022}).

\bibitem{peruzzo_variational_2014}
\bibinfo{author}{Peruzzo, A.} \emph{et~al.}
\newblock \bibinfo{title}{A variational eigenvalue solver on a photonic quantum
  processor}.
\newblock \emph{\bibinfo{journal}{Nature Communications}}
  \textbf{\bibinfo{volume}{5}} (\bibinfo{year}{2014}).

\bibitem{cerezo_variational_2021}
\bibinfo{author}{Cerezo, M.}, \bibinfo{author}{Arrasmith, A.},
  \bibinfo{author}{Babbush, R.} \& \bibinfo{author}{al, e.}
\newblock \bibinfo{title}{Variational quantum algorithms}.
\newblock \emph{\bibinfo{journal}{Nature Review Physics}}
  \textbf{\bibinfo{volume}{3}}, \bibinfo{pages}{625} (\bibinfo{year}{2021}).

\bibitem{preskill_quantum_2018}
\bibinfo{author}{Preskill, J.}
\newblock \bibinfo{title}{Quantum {Computing} in the {NISQ} era and beyond}.
\newblock \emph{\bibinfo{journal}{Quantum}} \textbf{\bibinfo{volume}{2}},
  \bibinfo{pages}{79} (\bibinfo{year}{2018}).

\bibitem{grover_fast_1996}
\bibinfo{author}{Grover, L.~K.}
\newblock \bibinfo{title}{A {Fast} {Quantum} {Mechanical} {Algorithm} for
  {Database} {Search}}.
\newblock In \emph{\bibinfo{booktitle}{Annual {Acm} {Symposium} on {Theory} of
  {Computing}}}, \bibinfo{pages}{212--219} (\bibinfo{publisher}{ACM},
  \bibinfo{year}{1996}).

\bibitem{zalka_grovers_1999}
\bibinfo{author}{Zalka, C.}
\newblock \bibinfo{title}{Grover's quantum searching algorithm is optimal}.
\newblock \emph{\bibinfo{journal}{Physical Review A}}
  \textbf{\bibinfo{volume}{60}}, \bibinfo{pages}{2746--2751}
  (\bibinfo{year}{1999}).

\bibitem{bennett_quantum_2022}
\bibinfo{author}{Bennett, T.} \& \bibinfo{author}{Wang, J.~B.}
\newblock \bibinfo{title}{Quantum optimisation via maximally amplified states}.
\newblock \bibinfo{howpublished}{Preprint at https://arxiv.org/abs/2111.00796}
  (\bibinfo{year}{2021}).

\bibitem{wurtz_maxcut_2021}
\bibinfo{author}{Wurtz, J.} \& \bibinfo{author}{Love, P.}
\newblock \bibinfo{title}{{MaxCut} quantum approximate optimization algorithm
  performance guarantees for p{\textgreater}1}.
\newblock \emph{\bibinfo{journal}{Physical Review A}}
  \textbf{\bibinfo{volume}{103}}, \bibinfo{pages}{042612}
  (\bibinfo{year}{2021}).

\bibitem{zhu_adaptive_2022}
\bibinfo{author}{Zhu, L.} \emph{et~al.}
\newblock \bibinfo{title}{Adaptive quantum approximate optimization algorithm
  for solving combinatorial problems on a quantum computer}.
\newblock \emph{\bibinfo{journal}{Physical Review Research}}
  \textbf{\bibinfo{volume}{4}}, \bibinfo{pages}{033029} (\bibinfo{year}{2022}).

\bibitem{nocedal_numerical_2006}
\bibinfo{author}{Nocedal, J.} \& \bibinfo{author}{Wright, S.~J.}
\newblock \emph{\bibinfo{title}{Numerical optimization}}.
\newblock Springer series in operations research
  (\bibinfo{publisher}{Springer}, \bibinfo{address}{New York},
  \bibinfo{year}{2006}), \bibinfo{edition}{2nd ed} edn.

\bibitem{nelder_simplex_1965}
\bibinfo{author}{Nelder, J.~A.} \& \bibinfo{author}{Mead, R.}
\newblock \bibinfo{title}{A {Simplex} {Method} for {Function} {Minimization}}.
\newblock \emph{\bibinfo{journal}{The Computer Journal}}
  \textbf{\bibinfo{volume}{7}}, \bibinfo{pages}{308--313}
  (\bibinfo{year}{1965}).

\bibitem{markowitz_portfolio_1952}
\bibinfo{author}{Markowitz, H.}
\newblock \bibinfo{title}{Portfolio {Selection}}.
\newblock \emph{\bibinfo{journal}{The Journal of Finance}}
  \textbf{\bibinfo{volume}{7}}, \bibinfo{pages}{77--91} (\bibinfo{year}{1952}).

\bibitem{slate_quantum_2021}
\bibinfo{author}{Slate, N.}, \bibinfo{author}{Matwiejew, E.},
  \bibinfo{author}{Marsh, S.} \& \bibinfo{author}{Wang, J.~B.}
\newblock \bibinfo{title}{Quantum walk-based portfolio optimisation}.
\newblock \emph{\bibinfo{journal}{Quantum}} \textbf{\bibinfo{volume}{5}},
  \bibinfo{pages}{513} (\bibinfo{year}{2021}).

\bibitem{jordan_fast_2005}
\bibinfo{author}{Jordan, S.~P.}
\newblock \bibinfo{title}{Fast quantum algorithm for numerical gradient
  estimation}.
\newblock \emph{\bibinfo{journal}{Physical review letters}}
  \textbf{\bibinfo{volume}{95}}, \bibinfo{pages}{050501}
  (\bibinfo{year}{2005}).

\bibitem{gilyen_optimizing_2019}
\bibinfo{author}{Gilyén, A.}, \bibinfo{author}{Arunachalam, S.} \&
  \bibinfo{author}{Wiebe, N.}
\newblock \bibinfo{title}{Optimizing quantum optimization algorithms via faster
  quantum gradient computation}.
\newblock In \emph{\bibinfo{booktitle}{Proceedings of the {Thirtieth} {Annual}
  {ACM}-{SIAM} {Symposium} on {Discrete} {Algorithms}}},
  \bibinfo{pages}{1425--1444} (\bibinfo{publisher}{SIAM},
  \bibinfo{year}{2019}).

\bibitem{chakrabarti_quantum_2020}
\bibinfo{author}{Chakrabarti, S.}, \bibinfo{author}{Childs, A.~M.},
  \bibinfo{author}{Li, T.} \& \bibinfo{author}{Wu, X.}
\newblock \bibinfo{title}{Quantum algorithms and lower bounds for convex
  optimization}.
\newblock \emph{\bibinfo{journal}{Quantum}} \textbf{\bibinfo{volume}{4}},
  \bibinfo{pages}{221} (\bibinfo{year}{2020}).

\bibitem{van_apeldoorn_convex_2020}
\bibinfo{author}{van Apeldoorn, J.}, \bibinfo{author}{Gilyén, A.},
  \bibinfo{author}{Gribling, S.} \& \bibinfo{author}{de~Wolf, R.}
\newblock \bibinfo{title}{Convex optimization using quantum oracles}.
\newblock \emph{\bibinfo{journal}{Quantum}} \textbf{\bibinfo{volume}{4}},
  \bibinfo{pages}{220} (\bibinfo{year}{2020}).

\bibitem{zhang_quantum_2021}
\bibinfo{author}{Zhang, C.}, \bibinfo{author}{Leng, J.} \& \bibinfo{author}{Li,
  T.}
\newblock \bibinfo{title}{Quantum algorithms for escaping from saddle points}.
\newblock \emph{\bibinfo{journal}{Quantum}} \textbf{\bibinfo{volume}{5}},
  \bibinfo{pages}{529} (\bibinfo{year}{2021}).

\bibitem{rebentrost_quantum_2019}
\bibinfo{author}{Rebentrost, P.}, \bibinfo{author}{Schuld, M.},
  \bibinfo{author}{Wossnig, L.}, \bibinfo{author}{Petruccione, F.} \&
  \bibinfo{author}{Lloyd, S.}
\newblock \bibinfo{title}{Quantum gradient descent and {Newton}’s method for
  constrained polynomial optimization}.
\newblock \emph{\bibinfo{journal}{New Journal of Physics}}
  \textbf{\bibinfo{volume}{21}}, \bibinfo{pages}{073023}
  (\bibinfo{year}{2019}).

\bibitem{kerenidis_quantum_2020}
\bibinfo{author}{Kerenidis, I.} \& \bibinfo{author}{Prakash, A.}
\newblock \bibinfo{title}{Quantum gradient descent for linear systems and least
  squares}.
\newblock \emph{\bibinfo{journal}{Physical Review A}}
  \textbf{\bibinfo{volume}{101}}, \bibinfo{pages}{022316}
  (\bibinfo{year}{2020}).

\bibitem{verdon_quantum_2019}
\bibinfo{author}{Verdon, G.}, \bibinfo{author}{Arrazola, J.~M.},
  \bibinfo{author}{Brádler, K.} \& \bibinfo{author}{Killoran, N.}
\newblock \bibinfo{title}{A quantum approximate optimization algorithm for
  continuous problems}.
\newblock \bibinfo{howpublished}{Preprint at https://arxiv.org/abs/1902.00409}
  (\bibinfo{year}{2019}).

\bibitem{enomoto_continuous-variable_2022}
\bibinfo{author}{Enomoto, Y.}, \bibinfo{author}{Anai, K.},
  \bibinfo{author}{Udagawa, K.} \& \bibinfo{author}{Takeda, S.}
\newblock \bibinfo{title}{Continuous-variable quantum approximate optimization
  on a programmable photonic quantum processor}.
\newblock \bibinfo{howpublished}{Preprint at https://arxiv.org/abs/2206.07214}
  (\bibinfo{year}{2022}).

\bibitem{QWbook2014}
\bibinfo{author}{Manouchehri, K.} \& \bibinfo{author}{Wang, J.~B.}
\newblock \emph{\bibinfo{title}{Physical implementation of quantum walks}}
  (\bibinfo{publisher}{Springer}, \bibinfo{year}{2014}).

\bibitem{qiang2016}
\bibinfo{author}{Qiang, X.} \emph{et~al.}
\newblock \bibinfo{title}{Efficient quantum walk on a quantum processor}.
\newblock \emph{\bibinfo{journal}{Nature Communications}}
  \textbf{\bibinfo{volume}{7}}, \bibinfo{pages}{11511} (\bibinfo{year}{2016}).

\bibitem{loke2017}
\bibinfo{author}{Loke, T.} \& \bibinfo{author}{Wang, J.~B.}
\newblock \bibinfo{title}{Efficient quantum circuits for continuous-time
  quantum walks on composite graphs}.
\newblock \emph{\bibinfo{journal}{Journal of Physics A: Mathematical and
  Theoretical}} \textbf{\bibinfo{volume}{50}}, \bibinfo{pages}{055303}
  (\bibinfo{year}{2017}).

\bibitem{zhou2017}
\bibinfo{author}{Zhou, S.~S.} \& \bibinfo{author}{Wang, J.~B.}
\newblock \bibinfo{title}{Efficient quantum circuits for dense circulant and
  circulant like operators}.
\newblock \emph{\bibinfo{journal}{Royal Society Open Science}}
  \textbf{\bibinfo{volume}{4}}, \bibinfo{pages}{160906} (\bibinfo{year}{2017}).

\bibitem{ostrouchov_parallel_1987}
\bibinfo{author}{Ostrouchov, G.}
\newblock \bibinfo{title}{Parallel computing on a hypercube: an overview of the
  architecture and some applications}.
\newblock In \emph{\bibinfo{booktitle}{19th Symposium on the Interface of
  Computer Science and Statistics}} (\bibinfo{year}{1987}).

\bibitem{verdon_universal_nodate}
\bibinfo{author}{Verdon, G.}, \bibinfo{author}{Pye, J.} \&
  \bibinfo{author}{Broughton, M.}
\newblock \bibinfo{title}{A universal training algorithm for quantum deep
  learning}.
\newblock \bibinfo{howpublished}{Preprint at https://arxiv.org/abs/1806.09729}
  (\bibinfo{year}{2018}).

\bibitem{chan_embedding_1991}
\bibinfo{author}{Chan, M.~Y.}
\newblock \bibinfo{title}{Embedding of {Grids} into {Optimal} {Hypercubes}}.
\newblock \emph{\bibinfo{journal}{SIAM Journal on Computing}}
  \textbf{\bibinfo{volume}{20}}, \bibinfo{pages}{834--864}
  (\bibinfo{year}{1991}).

\bibitem{hales_improved_2000}
\bibinfo{author}{Hales, L.} \& \bibinfo{author}{Hallgren, S.}
\newblock \bibinfo{title}{An improved quantum {Fourier} transform algorithm and
  applications}.
\newblock In \emph{\bibinfo{booktitle}{Proceedings 41st {Annual} {Symposium} on
  {Foundations} of {Computer} {Science}}}, \bibinfo{pages}{515--525}
  (\bibinfo{year}{2000}).

\bibitem{frigo_fastest_1997}
\bibinfo{author}{Frigo, M.} \& \bibinfo{author}{Johnson, S.~G.}
\newblock \bibinfo{title}{The {Fastest} {Fourier} {Transform} in the {West}:}
  (\bibinfo{year}{1997}).

\bibitem{frigo_design_2005}
\bibinfo{author}{Frigo, M.} \& \bibinfo{author}{Johnson, S.}
\newblock \bibinfo{title}{The {Design} and {Implementation} of {FFTW3}}.
\newblock \emph{\bibinfo{journal}{Proceedings of the IEEE}}
  \textbf{\bibinfo{volume}{93}}, \bibinfo{pages}{216--231}
  (\bibinfo{year}{2005}).

\bibitem{gao_implementing_2012}
\bibinfo{author}{Gao, F.} \& \bibinfo{author}{Han, L.}
\newblock \bibinfo{title}{Implementing the {Nelder}-{Mead} simplex algorithm
  with adaptive parameters}.
\newblock \emph{\bibinfo{journal}{Computational Optimization and Applications}}
  \textbf{\bibinfo{volume}{51}}, \bibinfo{pages}{259--277}
  (\bibinfo{year}{2012}).

\bibitem{quop_mpi}
\bibinfo{author}{Matwiejew, E.}
\newblock \bibinfo{title}{{QuOp}\_{MPI}: a {Python} module for parallel
  distributed memory simulation and design of {Quantum} {Variational}
  {Algorithms}.}
\newblock
  \bibinfo{howpublished}{\url{https://github.com/Edric-Matwiejew/QuOp_MPI}}
  (\bibinfo{year}{2022}).

\bibitem{galletly_evolutionary_1998}
\bibinfo{author}{Galletly, J.}
\newblock \bibinfo{title}{Evolutionary {Algorithms} in {Theory} and {Practice}:
  : {Evolution} {Strategies}, {Evolutionary} {Programming}, {Genetic}
  {Algorithms}}.
\newblock \emph{\bibinfo{journal}{Kybernetes}} \textbf{\bibinfo{volume}{27}},
  \bibinfo{pages}{979--980} (\bibinfo{year}{1998}).

\bibitem{haupt_practical_2004}
\bibinfo{author}{Haupt, R.~L.} \& \bibinfo{author}{Haupt, S.~E.}
\newblock \emph{\bibinfo{title}{Practical genetic algorithms with {CD}-{Rom}}}
  (\bibinfo{publisher}{J. Wiley}, \bibinfo{address}{New York},
  \bibinfo{year}{2004}), \bibinfo{edition}{2} edn.

\bibitem{vanaret_hybridization_2015}
\bibinfo{author}{Vanaret, C.}
\newblock \bibinfo{title}{Hybridization of interval methods and evolutionary
  algorithms for solving difficult optimization problems}.
\newblock \bibinfo{howpublished}{Preprint at https://arxiv.org/abs/2001.11465}
  (\bibinfo{year}{2020}).

\end{thebibliography}

\clearpage

\section{Appendix}

\renewcommand{\thesubsection}{\Alph{subsection}}

\titleformat{\subsection}
{\bfseries}
{\thesubsection. }{0in}{}

\subsection{Grover's Search and QVA Convergence}\label{app:grovers}

Grover's search is a deterministic quantum state amplification algorithm consisting of a pair of alternating unitaries that are applied to an equal superposition over $\ket{k}$~\cite{grover_fast_1996}. These unitaries are equivalent to the QVA phase-shift unitary (\cref{eq:phase-shift}) and the complete-graph QAOA mixing unitary (\cref{eq:qaoa_mixer}) with $t=\frac{\pi}{K}$ and $\gamma=\pi$ for all $p$~\cite{bennett_quantum_2022}.  In place of $\hat{Q}$ is an oracle that returns $1$ for the target (`marked') $\ket{k}$ and $0$ otherwise. 

For a single marked state, Grover's Search achieves complete (or near complete) convergence with $p \approx \pi \frac{\sqrt{K}}{4}$ iterations. For problems of practical interest, the $p$ required far exceeds the circuit-depth capabilities of NISQ devices. A variation on Grover's search, the restricted depth Grover's search (RDGS), carries out a fixed number of Grover iterations (less than $\frac{\pi \sqrt{K}}{4}$). The probability of measuring the marked state after $p$ iterations is,
\begin{equation}
	G(p, K) = \sin^2 \left[ \left(p + \frac{1}{2}\right) 2 \arcsin\left( \sqrt{\frac{1}{K}} \right) \right].
	\label{eq:grovers}
\end{equation}

Notably, the convergence given by \cref{eq:grovers} is proven to be optimal for an unstructured quantum search in the number of calls to the oracle or, equivalently, evaluations of $f(\bm{x}_k)$ in quantum parallel~\cite{zalka_grovers_1999}. As such, the RDGS provides a threshold for identifying efficient leveraging of solution space structure in QVAs.

\subsection{Optimisation Test-Functions}\label{app:test_functions}

Test-functions used for benchmarking of the QMOA, QOWE and the QAOA. Functions are listed with the function minimum and search domain. All functions are taken from \cite{galletly_evolutionary_1998,haupt_practical_2004}, unless otherwise noted.

{\small
\setlist[enumerate]{itemsep=0mm}
\begin{enumerate}
	\item Sphere function,	      
	      \[f(\boldsymbol{x}) = \sum_{d=0}^{D-1} x_{d}^{2} \]
	      
	      \noindent where $f_\text{min} = 0$ at $x_d = 0$ and $-2 \le x_{d} \le 2$.
	\item Rosenbrock function,      
	      \[ f(\boldsymbol{x}) = \sum_{d=0}^{D-2} \left[ 100 \left(x_{d+1} - x_{d}^{2}\right)^{2} + \left(1 - x_{d}\right)^{2}\right] \]
	      
	      \noindent where $f\text{min} = 0$ at $x_d = 1$ and $-3 \le x_{d} \le 3$.
	       
	\item Styblinski–Tang function,	      
	      \[ f(\boldsymbol{x}) = \frac{\sum_{d=0}^{D-1} x_{d}^{4} - 16x_{d}^{2} + 5x_{d}}{2} \]
	      
	      \noindent where $f_\text{min} = -39.16617 d$ at $x_d= -2.903534$ and $-5\le x_{d} \le 5$.
	          
	\item  Rastrigin function,	      
	      \[f(\mathbf{x}) = 10 D + \sum_{d=0}^{D-1} \left[x_d^2 - 10 \cos(2 \pi x_d)\right] \]
	      
	      \noindent where $f_\text{min} = 0$ at $x_d = 0$ and $-5.12\le x_{d} \le 5.12$.
	      
	\item Ackley function,	      
	      \begin{align*}
	      	\ f(x,y) = -20\exp\left[-0.2\sqrt{0.5\left(x^{2}+y^{2}\right)}\right] -\exp\left[0.5\left(\cos 2\pi x + \cos 2\pi y \right)\right] + e + 20 
	      \end{align*}
	          
	      \noindent where $f_\text{min} = 0$ at $(0,0)$ and $-5\le x,y \le 5$.
	      
	\item Beale function,	      
	      \begin{align*}
	      	f(x,y) = \left( 1.5 - x + xy \right)^{2} + \left( 2.25 - x + xy^{2}\right)^{2} \left(2.625 - x+ xy^{3}\right)^{2}                                      
	      \end{align*}
	      
	      \noindent where $f_\text{min} = 0$ at $(3, 0.5)$ and $-4.5 \le x,y \le 4.5$.
	          
	\item Goldstein–Price function,
	      \begin{align*}
	      	  f(x,y) & = \left[1+\left(x+y+1\right)^{2}\left(19-14x+3x^{2}-14y+6xy+3y^{2}\right)\right]     \\ & \qquad \times \left[30+\left(2x-3y\right)^{2}\left(18-32x+12x^{2}+48y-36xy+27y^{2}\right)\right] 
	      \end{align*}
	      
	      \noindent where $f_\text{min} = 3$ at $(0, -1)$ and $-2 \le x,y \le 2$.
	      
	\item Booth function,	      
	      \[ f(x,y) = \left( x + 2y -7\right)^{2} + \left(2x +y - 5\right)^{2} \]
	      
	      \noindent where $f_\text{min} = 0$ at $(1,3)$ and $-10 \le x,y \le 10$.
	      
	\item Bukin function N.6,	      
	      \[ f(x,y) = 100\sqrt{\left\vert y - 0.01x^{2}\right\vert} + 0.01 \left\vert x+10 \right\vert.\quad \]
	      
	      \noindent where $f_\text{min} = 0$ at $(-10,1)$ and $-15 \le x \le -5$ and $-3\le y \le 3$.
	      
	\item Matyas function,	      
	      \[ f(x,y) = 0.26 \left( x^{2} + y^{2}\right) - 0.48 xy \]
	      
	      \noindent where $f_\text{min} = 0$ at $(0,0)$ and $-10\le x,y \le 10$.
	          
	\item L\'evi function N.13,	      
	      \begin{align*}
	      	f(x,y) & = \sin^{2} 3\pi x + \left(x-1\right)^{2}\left(1+\sin^{2} 3\pi y\right)  +\left(y-1\right)^{2}\left(1+\sin^{2} 2\pi y\right)                    
	      \end{align*}
	      
	      \noindent where $f_\text{min} = 0$ at $(1,1)$ and $-10\le x,y \le 10$.
	      
	\item Himmelblau's function,	      
	      \[f(x, y) = (x^2+y-11)^2 + (x+y^2-7)^2.\]
	      
	      \noindent where $f_\text{min} = 0$ at $(3.0,  2.0)$, $(-2.805118, 3.131312)$, $(-3.779310, -3.283186)$ and $(3.584428, -1.848126)$, and $-5\le x,y \le 5$.
	      
	\item Three-hump camel function,	    
	      \[f(x,y) = 2x^{2} - 1.05x^{4} + \frac{x^{6}}{6} + xy + y^{2}\]
	      
	      \noindent where $f_\text{min} = 0$ at $(0,0)$ and $-5\le x,y \le 5$.
	      
	\item Easom function,	      
	      \[ f(x,y) = -\cos \left(x\right)\cos \left(y\right) \exp\left(-\left(\left(x-\pi\right)^{2} + \left(y-\pi\right)^{2}\right)\right) \]
	      
	      \noindent where $f_\text{min} = -1$ at $(\pi , \pi)$ and $-100\le x,y \le 100$.
	      
	\item Cross-in-tray function,	      
	      \begin{align*}
	      	f(x,y) & =                                                                                                                            -0.0001 \left[ \left\vert \sin x \sin y \exp \left(\left\vert100 - \frac{\sqrt{x^{2} + y^{2}}}{\pi} \right\vert\right)\right\vert + 1 \right]^{0.1} 
	      \end{align*}
	      
	      \noindent where $f_\text{min} = -2.06261$ at $(\pm 1.34941, \pm1.34941)$ and $-10\le x,y \le 10$.

	\item Eggholder function \cite{vanaret_hybridization_2015},	    
	      \begin{align*}
	      	f(x,y) & =                                                                                                                            - \left(y+47\right) \sin \sqrt{\left\vert\frac{x}{2}+\left(y+47\right)\right\vert} - x \sin \sqrt{\left\vert x - \left(y + 47 \right)\right\vert} 
	      \end{align*}
	         
	      \noindent where $f_\text{min} = -959.6407$ at $(512, 404.2319)$ and $-512\le x,y \le 512$.
	      
	\item H\"older table function,	      
	      \[f(x,y) = - \left\vert\sin x \cos y \exp \left(\left\vert1 - \frac{\sqrt{x^{2} + y^{2}}}{\pi} \right\vert\right)\right\vert\]
	      
	      \noindent where $f_\text{min} = -19.2085$ at $(\pm 8.05502, \pm 9.66459)$ and $-10\le x,y \le 10$.
	      
	\item McCormick function,      
	      \[f(x,y) = \sin \left(x+y\right) + \left(x-y\right)^{2} - 1.5x + 2.5y + 1 \]
	      
	      \noindent where $f_\text{min} = -1.9133$ at $(-0.54719,-1.54719)$ and, $-1.5\le x \le 4$ and $-3\le y \le 4$.
	      
	\item Schaffer function N. 2,      
	      \[f(x,y) = 0.5 + \frac{\sin^{2}\left(x^{2} - y^{2}\right) - 0.5}{\left[1 + 0.001\left(x^{2} + y^{2}\right) \right]^{2}} \]
	      
	      \noindent where $f_\text{min} = 0$ at $(0, 0)$ and $-100\le x,y \le 100$.
	      
	\item Schaffer function N. 4,	      
	      \[f(x,y) = 0.5 + \frac{\cos^{2}\left[\sin \left( \left\vert x^{2} - y^{2}\right\vert\right)\right] - 0.5}{\left[1 + 0.001\left(x^{2} + y^{2}\right) \right]^{2}} \]
	      
	      \noindent where $f_\text{min} = 0.292579$ at $(0, \pm 1.25313)$ and $(\pm 1.25313, 0)$, and $-100\le x,y \le 100$.
	          
\end{enumerate}
}

\clearpage

\subsection{QMOA Mixing Structure and Walk Parameterisation}\label{app:qmoa_simulation}

\vspace{-0.3cm}

\renewcommand{\thefigure}{C.\arabic{figure}}
\setcounter{figure}{0}

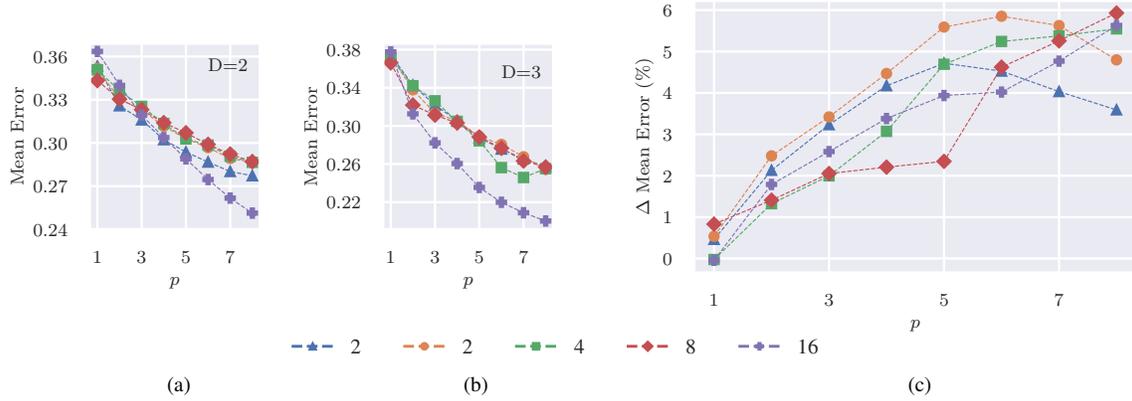
\begin{figure}[!ht]
		
	\centering
    \begin{tikzpicture}
    
    \node[inner sep=0pt] (A) at (0,0)
    {\includegraphics[width=0.23\linewidth]{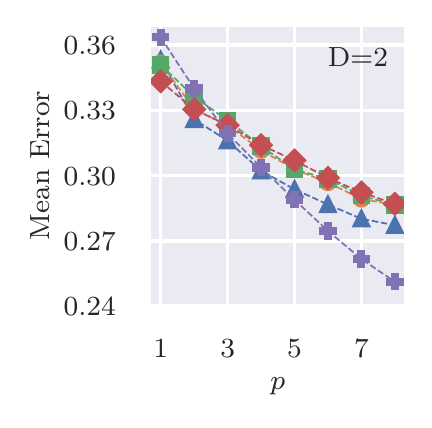}};
    \node[inner sep=0pt] (B) at (3.9,0)
    {\includegraphics[width=0.23\linewidth]{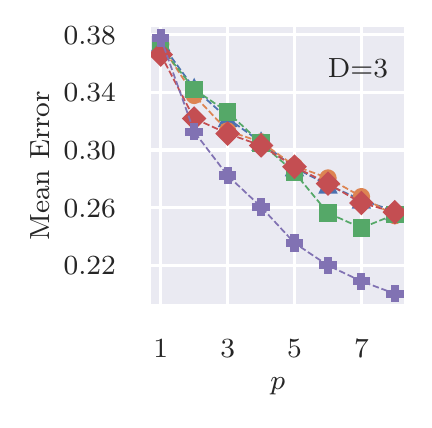}};
    \node[inner sep=0pt] (C) at (10,0)
    {\includegraphics[width=0.43\linewidth]{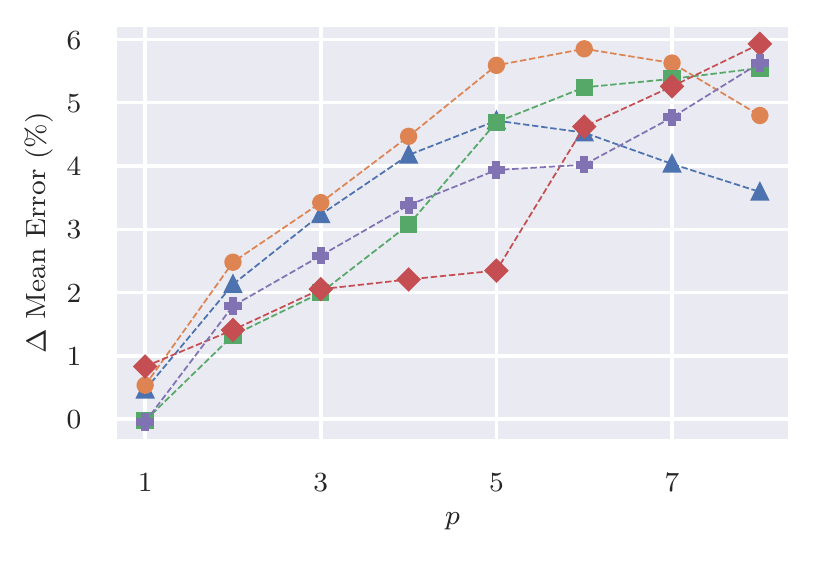}};
    
    \node[inner sep=0pt] (L) at (5.6,-2.4)
    {\includegraphics[width=0.46\linewidth]{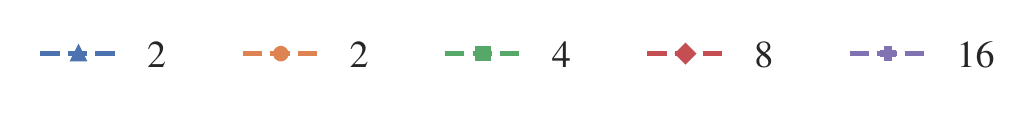}};
    
    \node[] (a) at (0.6, -2.9)  {\footnotesize(a)};
    \node[] (b) at (4.55, -2.9)  {\footnotesize(b)};
    \node[] (c) at (10.45, -2.9)  {\footnotesize(c)};
    \end{tikzpicture}		

    \vspace{-0.2cm}

	\caption{Mean error for the QMOA applied to the (a) $D=2$ and (b) $D=3$ test-functions with $N=32$ using $\hat{C}$ of varying vertex degree for $D=2$ ($10$ qubits) and $D=3$ ($15$ qubits). The presented mean error is the average over the respective test function sets. (c) The percentage difference in the QMOA mean error for the $D=2$ test function set with $N=32$ ($15$ qubits) for independently and non-independently parameterised $t_d$. Positive values are in favour of dimensionally independent $t_d$ parameterisation.}
	\label{fig:qmoa_tests}
		
\end{figure}

\vspace{-0.2cm}

The mean error of the QMOA was assessed for $\hat{C}$ of varying vertex degree on the test-functions for $D=2$ (all functions from App.~\ref{app:test_functions}) and $D=3$ (the Sphere, Ackley, Rosenbrock,  Styblinski-Tang, and Rastrigin functions). \cref{fig:qmoa_tests} (a) and (b) show the mean error averaged over the respective test-function sets. The different $\hat{C}$ achieve a similar mean error up to $p=5$ for $D=2$ and $p=2$ for $D=3$. At higher $p$, a complete graph $\hat{C}$ (vertex degree $31$) has the lowest mean error, and the difference between it and the next lowest mean error increases with $p$. Optimisation of the $D=2$ test set was repeated with $t_d$ for each dimension parameterised by the same value at each $p$. In \cref{fig:qmoa_tests} (c) we see that dimensionally independent $t_d$ parameterisation is advantageous at $p > 1$ for all of the considered $\hat{C}$. For $\hat{C}$ with vertex degrees of $16$ and $31$ the difference increases over the range of considered $p$.

\vspace{-0.2cm}

\subsection{Fitting Curves for QMOA (Complete) and QAOA (Hypercube) Scaling} \label{app:fitting}

\vspace{-0.4cm}

\renewcommand{\thefigure}{D.\arabic{figure}}
\setcounter{figure}{0}

\begin{figure}[h!]
	\centering

	\includegraphics[width=0.32\columnwidth]{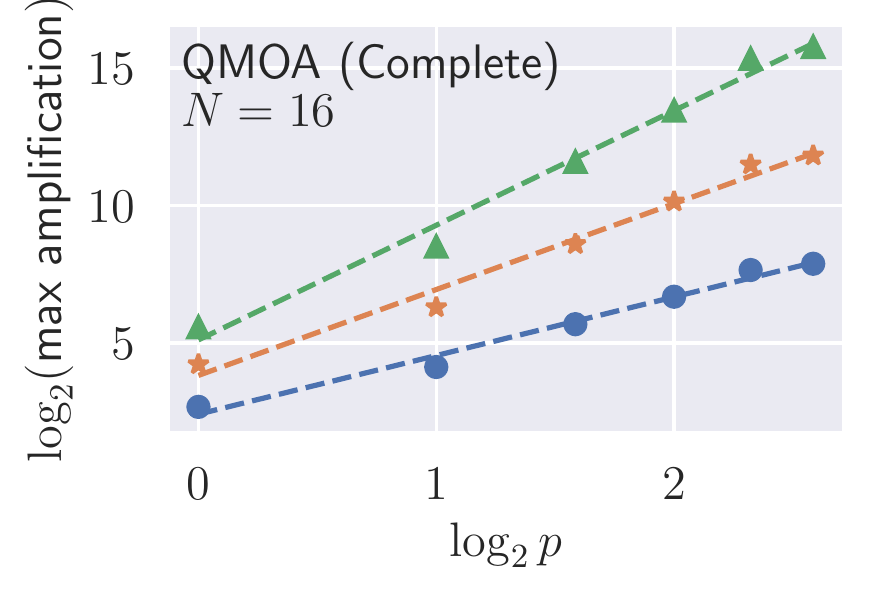}
	\includegraphics[width=0.32\columnwidth]{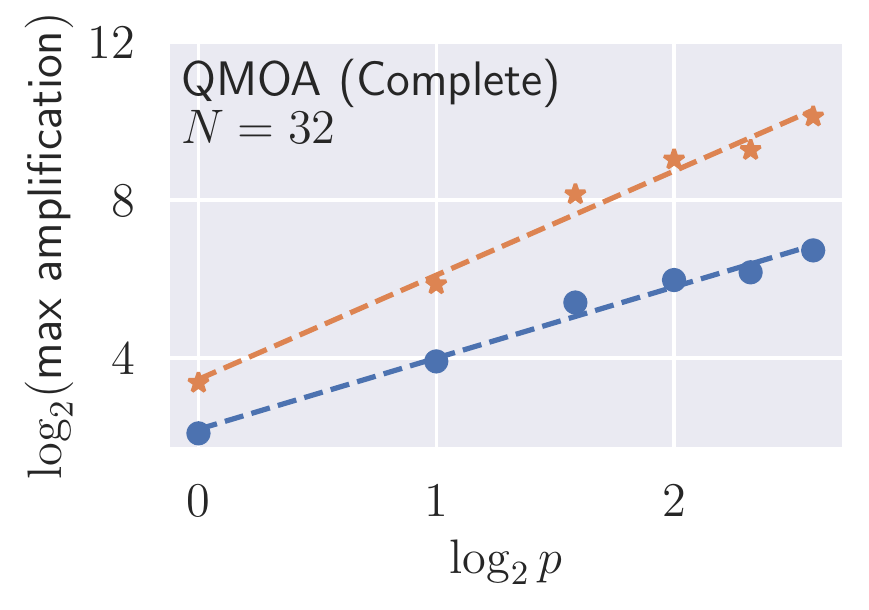}
	\includegraphics[width=0.32\columnwidth]{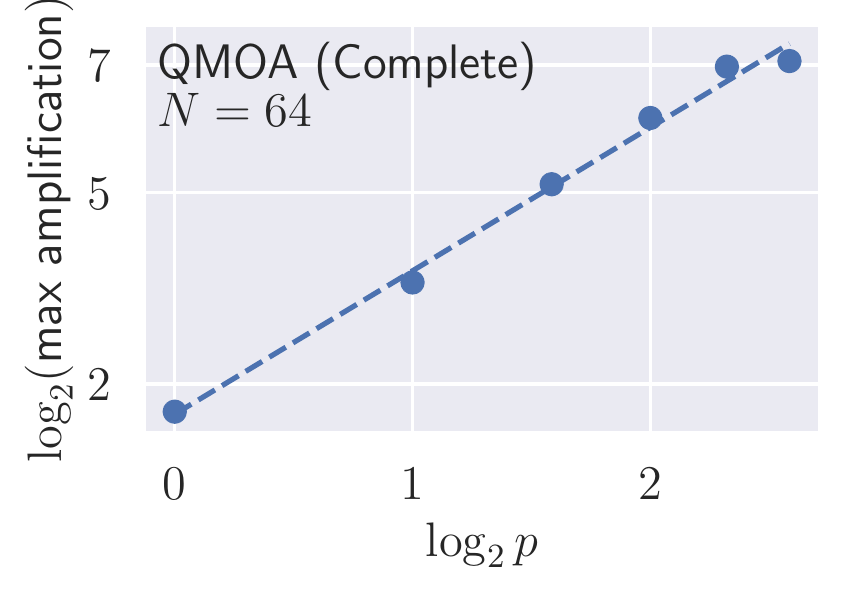}

	\includegraphics[width=0.32\columnwidth]{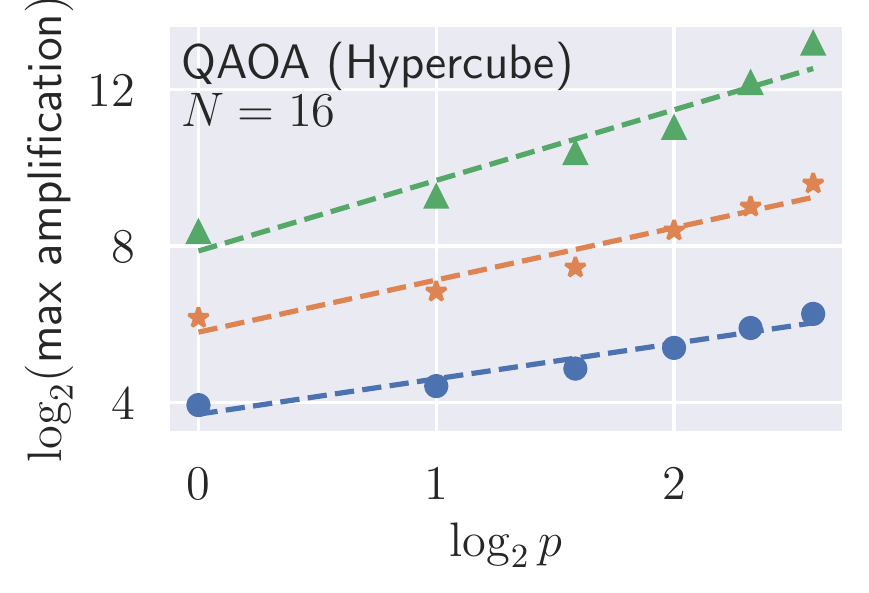}
	\includegraphics[width=0.32\columnwidth]{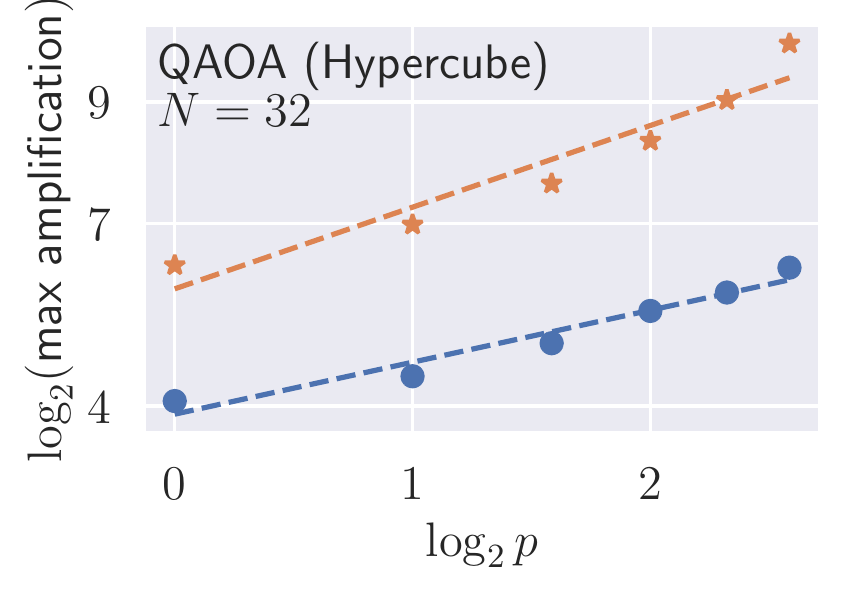}
	\includegraphics[width=0.32\columnwidth]{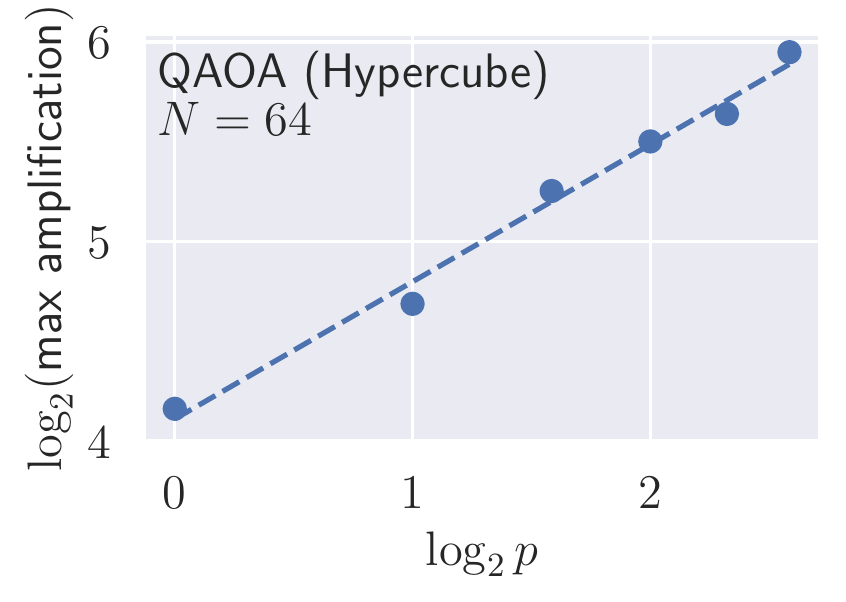}	

    \vspace{-0.2cm}

	\includegraphics[width=0.35\columnwidth]{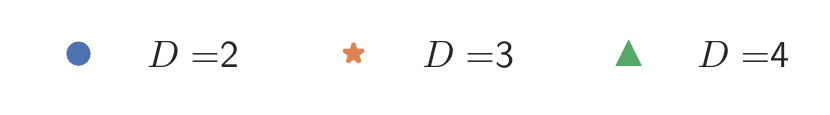}
	
	\vspace{-0.5cm}
	
    \caption{The maximum amplification for QMOA (Complete) and QAOA (Hypercube) optimisation of the RF from $p=1$ to $6$ with $N=32$ to $64$ in $D=2$ to $4$. The dotted lines show the line of best fit for $\log_2 (\text{maximum amplification}) = \log_2 C + \alpha D \log_2 p$. Each point is the mean of $60$ simulation repeats.}
		
    \label{fig:fitting}
	
\end{figure}

\FloatBarrier

\subsection{Global Minimisation with the QMOA and Nelder-Mead} \label{app:sampling}

The QMOA was applied to the context of classically assisted optimisation of the STF and RF. Variational parameters of the QMOA were optimised according to  $\E{Q}$ estimated from sample sets drawn according to $\vert \E{k \, \vert \, \bm{t}, \bm{\gamma}} \vert^2$. The previously referenced implementation of the Nelder-Mead algorithm was used at its default settings for optimisation of $(\bm{t}, \bm{\gamma})$, and a sample size of $30$ was found to be a good balance between classical optimisation convergence and the number of $f(\bm{x}_k)$ evaluations. The minimum $\bm{x}_k$ from each sample set was used as the starting point for further classical optimisation. Figure \ref{fig:assisted_sampling} shows the speedup achieved in terms of $f$ evaluations in identification of (the non-discretised) ${\bm{x}^*}$ compared to repeated application of the Nelder-Mead algorithm alone with $\epsilon = 10^{-4}$. Ansatz depth, $p$, was $2$ for the STF and $5$ for the RF. Speedup is reported as a ratio of the number of $f$ evaluations,
\[\text{speedup} = \frac{\text{fev}_\text{Nelder-Mead}}{\text{fev}_\text{assisted}} \]
which, for the QMOA with the Nelder-Mead algorithm, was $\text{fev}_\text{assisted} = 30(p+1) \text{fev}_\text{QMOA} + \text{fev}_\text{Nelder-Mead}$ where $\text{fev}_\text{QMOA}$ is the number of $\E{Q}$ estimations and $\text{fev}_\text{Nelder-Mead}$ is the total $f$ evaluations over $\text{fev}_\text{QMOA}$ repeats of the Nelder-Mead algorithm.

\renewcommand{\thefigure}{E.\arabic{figure}}
\setcounter{figure}{0}
\begin{figure}[h!]
    \centering
    
	\includegraphics[width=0.49\linewidth]{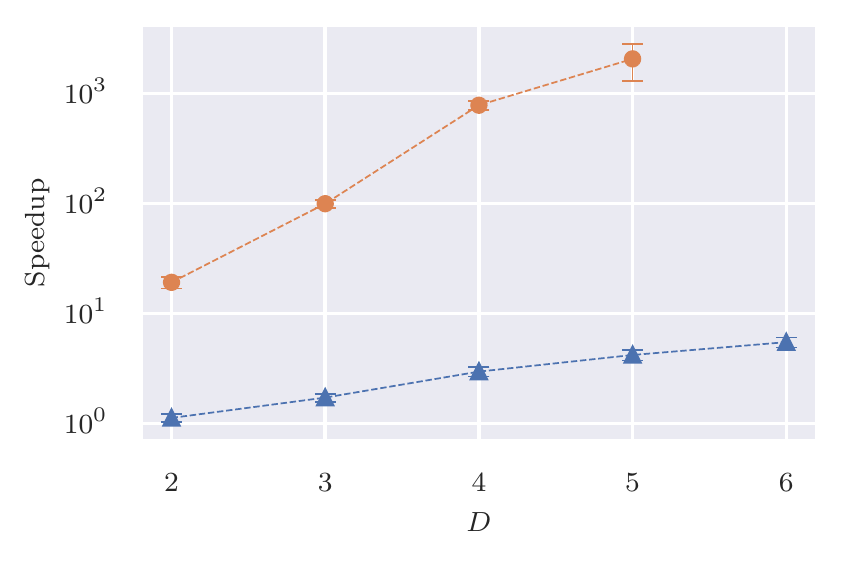}
	
	\includegraphics[width=0.55\linewidth]{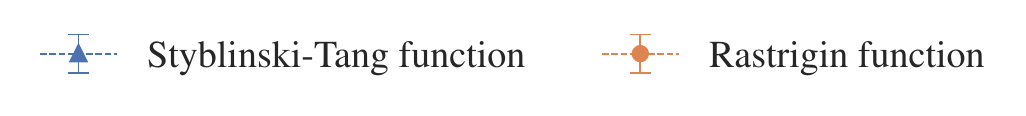}
	
	\caption{Speedup in identifying $\min(f_k)$ with $\epsilon = 10^{-4}$ for the STF and RF using the QMOA with the Nelder-Mead algorithm over Nelder-Mead alone. The QMOA optimised with $N=16$ grid points per $D$ for a total of $8$ qubits at $D=2$, $12$ qubits at $D=3$, $16$ qubits at $D=4$, $20$ qubits at $D=5$, and $24$ qubits at $D=6$. Each point depicts the mean and sample standard deviation of $200$ repeats.}
		
	\label{fig:assisted_sampling}
\end{figure}

For both functions, the quantum-assisted optimisation provided a speedup, with a maximum speedup of $5.48$ at $D=6$ for the STF and $2062$ at $D=5$ for the RF. Again, these results demonstrate that the QMOA algorithm provides the best advantage for oscillatory and high-dimensional functions. At $D=6$, quantum-assisted optimisation identified $\bm{x}^*$ of the RF in an average of $15 \times 10^5 \pm 2\times 10^4$ evaluations. In comparison, Nelder-Mead alone failed to converge to $\bm{x}^*$ over $3000$ repeats. The low ansatz depth required to provide speedup ($p=2$ for the STF and $p=5$ for the RF) is consistent with the scaling behaviour described observed in \cref{fig:continuous_depth_plots}. 

\end{document}